\documentclass[preprint]{aastex}

\shorttitle{Eccentricity Distribution of Gas Giants}

\shortauthors{Ida et al.}

\begin{document}

\title{Toward a Deterministic Model of Planetary Formation VII: \\
Eccentricity Distribution of Gas Giants}

\author{S. Ida$^1$, D. N. C. Lin$^{2,3}$, and M. Nagasawa$^4$}
\affil{1) Earth-Life Science Institute, Tokyo Institute of Technology,
Meguro-ku, Tokyo 152-8550, Japan \\
2) UCO/Lick Observatory, University of California, 
Santa Cruz, CA 95064, USA \\
3) Kavli Institute for Astronomy and Astrophysics, Peking 
University, Beijing, China\\
4) Department of Earth and Planetary Sciences, Tokyo Institute of Technology,
Meguro-ku, Tokyo 152-8551, Japan}
\email{ida@geo.titech.ac.jp,lin@ucolick.org,nagasawa.m.ad@m.titech.ac.jp}

\begin{abstract}
The ubiquity of planets and diversity of planetary systems reveal 
planet formation encompass many complex and competing processes.  
In this series of papers, we develop and upgrade a population 
synthesis model as a tool to identify the dominant physical effects 
and to calibrate the range of physical conditions.  Recent planet
searches leads to the discovery of many multiple-planet systems.
Any theoretical models of their origins must take into account
dynamical interaction between emerging protoplanets.  Here, we
introduce a prescription to approximate the close encounters between
multiple planets. We apply this method to simulate the growth, 
migration, and dynamical interaction of planetary systems. Our models
show that in relatively massive disks, several gas giants and 
rocky/icy planets emerge, migrate, and undergo dynamical instability.
Secular perturbation between planets leads to orbital crossings, eccentricity 
excitation, and planetary ejection.  In disks with modest masses,
two or less gas giants form with multiple super-Earths.  
Orbital stability in these systems is generally maintained and they retain
the kinematic structure after gas in their natal disks is depleted.  
These results reproduce the observed planetary 
mass-eccentricity and semimajor axis-eccentricity correlations.
They also suggest that emerging gas giants can scatter residual cores
to the outer disk regions. Subsequent {\it in situ} gas accretion 
onto these cores can lead to the formation of distant ($\ga 30$AU) 
gas giants with nearly circular orbits.
\end{abstract}

\section{Introduction}
The accelerating pace of observational discovery (with radial velocity, 
transit, microlensing, and direct imaging surveys) provide rich data sets 
on the statistical properties of extra solar planets and planetary 
systems.  In order to extract, from these data, useful information 
on the dominant physical processes and appropriate range of physical 
condition associated with planetary formation and evolution, 
population synthesis models have been developed. In a series of 
papers \citep[] [referred to as Paper I-V]{IL04a,IL04b,IL05,IL08a,IL08b},
we and others \citep{Mordasini09,Mordasini09b,Alibert11} have outlined
prescriptions for various physical processes and described attempts to
reproduce observational data with simulated models. Some of the relevant 
effects have been taken into account so far include the coagulation of 
planetesimals into embryos, gas accretion onto sufficiently massive 
cores, type I and II migration of individual protoplanets due to their 
interaction with their natal disks. 

Recent transit search by the Kepler space telescope indicates that 
many, if not most, super-Earths have siblings, i.e., they are members 
of multiple-planet systems. In these systems, gravitational 
interaction between planets affects not only the formation but 
also their evolution and destiny.  There have been several attempts 
to describe the dynamical evolution of these systems with N-body
simulations.  Due to the time consuming nature of such approaches,
only limited range of initial parameters have been explored. It is 
not clear whether these models adequately represent the distribution
of multiple planet systems.  

The population synthesis method is based on a statistical mechanics 
approach to evaluate the most likely outcome of dynamical interactions 
between planets. In order to describe the statistical outcome of 
gravitational interactions and collisions between rocky planetary 
embryos, we have upgraded our population synthesis method to generate 
a series of many simulations.  In paper VI \citep{IL10}, we described
our analytic prescription for the collisions and close scatterings between 
modest-mass planets.  We applied this method to show that 
super-Earths may be the {\it in situ} merger products of a 
population of embryos with Earth-mass or less which converged towards the
proximity of their host stars through type I migration.

Radial velocity survey show that modest to long-period 
($>7-10$ days) gas giants also have significant eccentricity.  
These planets most likely have formed in their natal disks 
with nearly circular orbits as in the case of Jupiter and Saturn
(Pollack et al 1996, Ida \& Lin 2004).  Their eccentricity 
may have been excited by the gravitational perturbation of 
their planetary siblings (Rasio \& Ford 1996, Zhou et al 2007,
Jaric \& Tremaine 2008, Chatterjie et al 2008).  Indeed, many 
gas giant planets are members of known multiple-planet systems. 

In order to simulate the outcome of dynamical interaction 
between gas giants and their siblings including other gas giants 
as well as less-massive protoplanetary embryos, we construct in 
this paper, construct a "multiple-planet-in-a-disk model".  This 
approach is a natural continuation of the method presented in 
Paper VI. The main technical challenge is that the gravitational 
perturbations from gas giants significantly affect 
orbital configuration of the entire planetary system.

The method we have developed to approximate 
the evolution of systems of N gas giants is presented in \S2.
This prescription takes into account gravitational interactions 
between all the planets around their common host stars. 
It reproduces the general results of direct N-body 
simulations with a large reduction (by many orders of 
magnitude) in the computational cost.  A detailed description
of the method is given in the Appendix.

We incorporate this prescription into the latest version of 
the population synthesis models. In \S3, we briefly recapitulate
the methodology used in our population synthesis models including
prescriptions for: 1) disk structure and evolution, 2) planetesimal
growth and the planetary embryos' dynamical isolation, 3) migration
due to tidal interaction between protoplanets and their natal disks, 
and 4) resonant interaction between protoplanets.   

In \S4, we apply these prescription to simulate some sample 
multiple-planet systems.  These results show that that 1) 
the most favorable location for gas giant formation in the 
core-accretion scenario is at a few AU's, 2) the emergence
of gas giants leads to the scattering of residual protoplanetary
embryos, some of which may subsequently undergo gas accretion
and form later generation gas giants, 3) as a consequence of 
dynamical instability, some gas giants in multiple-planet
systems may be scattered to attain the observed eccentricity distribution.

An advantage of the population synthesis model is its 
computational efficiency.  
With it, we are able simulate in \S5 rich sets of planetary
systems based on the observed range of disk properties. 
We compare the results obtained from the population synthesis
models with the observational data.  
Our models reproduce the observed planetary mass - eccentricity 
and semimajor axis - eccentricity correlations.
These comparisons 
are important not only to calibrate some uncertainties 
in the model parameters but also to highlight the dominant
physical processes.

Finally in \S6, we summarize our results and discuss 
their implications.
               
\section{Modeling of eccentricity excitation and
ejection of giant planets as a result of gravitational instability}

The objective of this section is to introduce a simple-to-use and 
computationally-efficient prescription to approximate the dynamical 
interaction between multiple planets around common host stars.

\subsection{An overview of our technical approach}
Before outlining the technical details of our prescription, we 
first qualitatively describe the principles of our analytic 
approximation for the gravitational interactions between 
several gas giants. This approach is guided by the results 
generated from various N-body simulations of the dynamical 
interaction between giant planets.  Many such simulations 
have been carried out for the purpose of studying the 
statistical properties associated with close encounters 
between multiple planets especially with regard to the origin
of the eccentric as well as short-period Jupiter-mass planets 
\citep[e.g.][]{Rasio96, Weiden96, Ford01, Zhou07, 
FR08, Juric08, Marzari02, Nagasawa08, Chatterjee08}.  

Most of the N-body experiments have been carried out with 
idealized initial conditions. They show that 1) the onset 
of dynamical instability is sensitively determined by the 
planets' initial normalized (in terms of their Hill's radius) 
separation and 2) dynamically unstable systems undergo orbit 
crossing and relaxation which generally lead to a Rayleigh 
distribution in the planets' asymptotic eccentricity 
\citep{Ida_Makino92,Zhou07}.  
In order to check the validity of our prescription (to be 
presented below), we perform calculations with our analytic
prescriptions for the same initial conditions as those adopted 
by the previous N-body simulations.  

There are some analytic approximation for the stability of
planetary systems.  Two planets with masses $m_1$ and $m_2$ 
on initially circular orbits around a common host star with 
a mass $M_*$ ($ \gg m_1, m_2$) would immediately cross 
each other's path if the difference between their semimajor 
axes ($\Delta a$) is smaller than a critical value, 
\begin{equation}
\Delta a_{\rm c}  \simeq 2\sqrt{3} r_{\rm H}
\label{eq:deltaac}
\end{equation}
where $a$ is the average of their semimajor axis and
\begin{equation}
r_{\rm H} = ((m_1+m_2)/3 M_{\ast})^{1/3} a
\end{equation}
is there mutual Hill's radius.
But, orbital crossings in such a system would not occur if 
initially $\Delta a$ is slightly larger than $\Delta a_{\rm c}$  
\citep{Gladman93}.  In \S\ref{sec:twogiants}, we present 
a prescription to evaluate the asymptotic semimajor axis 
and eccentricity distributions of two-planet systems which
undergo orbital crossings.

In systems with three or more gas giants, the orbit-crossing
condition is significantly different. For a finite duration 
of time, systems with initial $\Delta a$ larger than 
$\Delta a_{\rm c}$ may be maintained in a quasi-stable state 
with a limited variation in the amplitude of eccentricity.  
However, in due course, such systems may undergo a transition 
with a rapid eccentricity increase which is followed by 
orbital crossings \citep{Chambers96, Lin_Ida97, Marzari02, Zhou07}. 
The time scale $\tau_{\rm cross}$ for the onset of this 
transition depends sensitively on the normalized initial 
orbital separation $\Delta a/r_{\rm H}$, the planet-to-star 
mass ratio, and weakly on the number of planets.

In the population synthesis model (see \S3), we consider 
the possibility that the giant planets and rocky/icy 
planetary embryos may form, coexist, and evolve 
contemporaneously in a gaseous environment.  During 
advanced stages of their formation, gas giants grow 
through runaway gas accretion, i.e., the time scale 
$\tau_{\rm KH}$ for them to double their mass $M_p$ 
decreases with $M_p$. Their Hill's radius and the 
width of their feeding zone also increases with their 
$M_p$. The eccentricity of the nearby planetary 
siblings and residual embryos (with $\Delta a 
< \Delta a_{\rm c}$) would be excited over their 
synodic periods $\tau_{\rm syn}$ with respect to the
orbits of the gas giants.

While multiple planets' eccentricity is excited by their 
secular interaction with each other, it is also damped
by their tidal interaction with their natal disks
\citep{Artymowicz93,Ward93,Tanaka04}. It is appropriate 
to take into account these competing effects \citep{Iwasaki02}
(see \S 3). Planet-disk interaction also leads to type I 
and II migration (for modest-mass embryos and high-mass 
gas giants respectively). Idealized prescriptions for 
isolated planets' migration have already been incorporated 
into early versions of population synthesis models (Papers I 
\& V). In a subsequent paper, we will develop and apply 
a new prescriptions for embryos' type I migration 
\citep{Paardekooper11} and consider the feedback on 
the disk structure by  multiple gas giants and their 
interference on each other's direction and speed of 
migration.

In this paper, we focus on dynamical interaction between 
multiple gas giants in the limit that the difference between 
their semimajor axis may be reduced during the course of 
their migration. Two gas giants would capture each 
other onto their mutual low-order mean-motion resonances 
provided the convergent speed is slow.  This effect
has already been analyzed with our population synthesis models
for interaction between rocky and icy embryos (Paper VI). 
In \S\ref{sec:twogiants}, we modify our prescription
to take into the nonlinear interaction and determine the
resonant capture condition between multiple gas giants.  
We also consider the critical converging speed above which the 
gas giants may overcome the resonances' barrier and intrude
into each other's Hill's sphere and undergo close encounters 
(see \S 3.6).

The intensity of planet-disk interaction reduces with the 
depletion of the disk gas.  Observational signatures of 
protostellar disks diminish on a time scale of 3-5 Myr.  
The accretion rate from the disk to the central stars 
(inferred from the UV veiling) also declines on a similar 
time scale \citep{Hartmann98}.  Our population synthesis 
models take into account disk evolution and adjust the 
effects of gas accretion, eccentricity damping, and orbital
migration accordingly.  After the disk gas is severely depleted, 
both gas damping and accretion cease while secular interaction 
between the planets persists.  

In Paper VI, we presented prescriptions for eccentricity 
excitations and merging events (giant impacts) among the
rocky/icy planetary embryos. For these relatively low-mass
objects, we construct an analytic approximation to determine
the outcome of the dynamical interaction between planets 
based on celestial mechanics. This prescription 
quantitatively reproduce the results of N-body simulations. 
We applied this prescription to simulate the evolution of 
systems of 10--20 embryos with a total mass of a 
few $M_{\oplus}$. This initial condition corresponds to the 
the advanced stage of oligarchic growth when the embryos 
become dynamically isolated. We showed that close encounters
generally lead to a velocity dispersion $\sigma$ which is 
a significant fraction of the embryos' surface escape speed. 
Within a few AU's, embryos are unlikely to be ejected 
because $\sigma$ is smaller than the local Keplerian speed 
$v_{\rm K}$.  After the gas depletion, repeated close encounters 
lead to mergers and the emergence of one or two Earth-mass 
planets with few residual embryos.  Thereafter, the magnitude of 
$\tau_{\rm cross}$ increases beyond the main sequence life 
span of solar type stars (ie $10^9-10^{10}$ yrs). We also 
used this prescription to show that some super-Earths may 
be the merger products of a population of embryos 
which migrated to the proximity of their host stars.  

In \S\ref{sec:threegiants} of this paper, we generalize our 
prescription to simulate the onset of dynamical instability 
and resultant orbital crossings among multiple gas giants in 
the outer regions of an evolving disk environment. The 
recoil motion excites the eccentricity of the retained 
planets to establish a Rayleigh's distribution similar 
to that observed among eccentric gas giants \citep{Rasio96, 
Weiden96, Lin_Ida97, Marzari02}.  In contrast to the 
embryos close to their host stars, the surface escape 
speed of typical gas giants exceeds the Keplerian speed
in the outer regions of their natal disks where they 
emerge.  Close encounters with gas giants at few AU 
from the host stars may lead to ejection rather 
than merger events. 

The crossing timescale $\tau_{\rm cross}$ increases with 
the semimajor axis separations and decreases with the 
orbital eccentricities. As the population of gas giants 
declines through merger and ejection events or migration 
and stellar consumption, their normalized semimajor 
axis separation ($\Delta a/r_{\rm H}$) is enlarged. 
Energy dissipation during cohesive collisions also
leads to eccentricity damping. These combined effects
lengthen $\tau_{\rm cross}$ by several orders of 
magnitude after each merger or ejection event.

Recent numerical simulations show that there is a $\sim 30\%$ 
chance for orbital crossings to eventually excite one of the gas
giants to attain nearly parabolic orbit by Kozai effect 
\citep{Nagasawa08}.\footnote{
The frequency of close
approaches to a host star during orbital crossings 
among planets found by \citet{Nagasawa08}
was confirmed by \citet{Beauge12}.
However, since \citet{Beauge12} used a formula of
weaker tidal damping, they did not find 
as many circularized close-in giants as \citet{Nagasawa08} found.}
The orbit of an inwardly scattered 
gas giant may be circularized if its pericenter distance 
is sufficiently close to the host star \citep{Ivanov}.  This 
scenario has been suggested as one of the mechanism for the 
origin of ``hot Jupiters'' \citep{Rasio96,Nagasawa08}. We have 
not incorporated tidal circularization and secular perturbations
including the Kozai effect into the current version of the 
population synthesis models, although the overall effects of 
secular eccentricity excitation is implicitly taken into 
account through the assumption that the orbital crossings 
start on timescale $\tau_{\rm cross}$.

\subsection{Two gas giants case}
\label{sec:twogiants}
The above description qualitatively outlines the overall
methodology of our population synthesis model. We now provide 
some technical details on its latest upgrade, i.e., the 
construction of an analytic plus Monte Carlo prescription.

We begin with the simplest case of dynamical interaction between
two gas giants. As we have indicated above, in this case, orbit 
crossing are expected to occur between two planets with nearly 
circular co-planar orbits which are initially separated in 
semimajor axis by $\Delta a < \Delta a_{\rm c}$. Extensive 
numerical simulations show that close encounters between 
these planets generally leads to the expansion of their semimajor 
axis separation and eccentricity excitation.  We construct 
a method to compute the outcome of these close encounters.  
This prescription is calibrated with the statistical results 
generated from numerical simulations \citep{Ford01,FR08}.  
Detailed prescriptions are described in the Appendix A1.

The procedure of our model for two-planet encounters 
is structured in the following manner.  
\begin{description}
\item[A1)] We first identify two interacting gas giants,
with mass $m_1$ and $m_2$. 
\item[A2)] We compute the maximum relative eccentricity 
attainable during a close scattering in terms of the 
two-body surface escape velocity ($v_{{\rm esc},12}$) scaled by 
the local Keplerian velocity ($v_{\rm K}$), such that, 
\begin{equation}
e_{{\rm esc},12} = \frac{v_{{\rm esc},12}}{v_{\rm K}}
\simeq 1.6 \left(\frac{m_1+m_2}{M_{\rm J}}\right)^{1/3}
  \left(\frac{\rho}{1{\rm gcm}^{-3}}\right)^{1/6}
  \left(\frac{a}{1{\rm AU}}\right)^{1/2}.
\label{eq:e_escG0}
\end{equation}
A probable eccentricity is generated for  
individual bodies $j (j=1,2)$ with
\begin{equation}
e_j^{\rm max} = \left( \frac{m_k {\cal R}_j}{m_1 + m_2} \right)
e_{\rm esc,12},
\label{eq:e_escj00}
\end{equation}
where $k \ne j$ refers to the perturbing companios and 
\begin{equation}
{\cal R}_j (x, \sigma) =(x/\sigma^2) {\rm exp} 
(-x^2/ 2 \sigma^2)
\end{equation} 
is a random number with a Rayleigh distribution of unit root mean 
square dispersion $\sigma$,
\item[A3-a)] If, at least one of the trial eccentricities
(given by Eq.~\ref{eq:e_escj00}) exceeds unity, the body with the 
larger eccentricity would be identified as a planet to be ejected.
The remaining planet is assumed to be retained. For it, a new
eccentricity is generated (with Eqs. \ref{eq:e_escjr20} 
and \ref{eq:ang_momentum20}) together with an associated 
semimajor axis (with Eq.~\ref{eq:energy_consv})
(see \S\ref{sec:twoeject}).
\item [A3-b)] If the trial eccentricities of both planets 
are less than unity, they would be adopted as the expected 
values for the two planets after their close encounters
with some damping (see step 3-b in Appendix A1). The 
associated semimajor axis changes are evaluated in accordance
with the prescription in \S\ref{sec:tworetain}.
\end{description}

This procedure is applied throughout the disk evolution.  
Prior to severe gas depletion, the damping of eccentricity  
due to the planets' tidal interaction with their natal disks 
is taken into account. The combined effect of planet-planet 
scattering and planet-disk tidal interaction induce the two planets
to enlarge their semimajor axis separation $\Delta a$ until they
become dynamically isolated with $\Delta a > \Delta a_{\rm c}$.  
The efficiency of gas damping declines during gas depletion. 
In the absence of any significant amount of residual disk gas, 
two planets with $\Delta a > \Delta a_{\rm c}$ would continue to 
scatter each other until either they merge with each other or 
one of them is ejected.

\subsubsection{Planetary escape after close encounters between 
two planets}
\label{sec:twoeject}

We consider the possibility that one member of a planet pair may attain sufficient 
recoil velocity to escape from the gravitational potential of their host 
star. We compare the results generated with our analytic plus Monte Carlo 
prescription with those obtained from N-body simulations which were 
computed without taking into account of any eccentricity damping effect
(due to either planet-disk interaction or dissipative collisions.).

Since both interacting planets were initially bound to their host stars, 
conservation of energy implies that at least one of them must be retained.
We label the retained and ejected planets as body 1 and 2 respectively. 
In Appendix A1, we derived the differential distribution function of 
the post-scattering eccentricity of the retained body 1 as
\begin{equation}
f(e)de \propto \frac{1}{\sqrt{2\pi} (e^*/3)} \exp \left(
-\frac{(e - e^*)^2}{2(e^*/3)^2} \right) de,
\label{eq:e_escjr20}
\end{equation} 
where
\begin{eqnarray}
e^* = 0.83 \frac{m_2}{m_1} - 0.17 \left(\frac{m_2}{m_1}\right)^2,
\label{eq:ang_momentum20}
\end{eqnarray}
and $m_1$ and $m_2$ are masses of retained and ejected planets 
respectively.  Since it takes less energy to eject the lighter 
body, we assume $m_1 > m_2$ which is also consistent with the 
results of N-body simulations. 

In Eqs.~(\ref{eq:e_escjr20}) and (\ref{eq:ang_momentum20}),
the eccentricity distribution function depends on the 
ratio rather than the individual values of 
$m_1$ and $m_2$.  The results of numerical simulation 
of close encounters between two equal mass planets 
\citet{FR08} also show a negligible dependence of 
$\mu = m_1/M_* = m_2/M_*$ (in the range between $0.001
-0.01$). Note from Eq.~(\ref{eq:e_escG0}), ejection cannot 
be produced from close encounters between planets with 
$\mu \ll 10^{-4} (\rho/1{\rm gcm}^{-3})^{-1/2} 
(a/1{\rm AU})^{-3/2}$).

\begin{figure}[btp]
  \epsscale{1.0}       
 \plotone{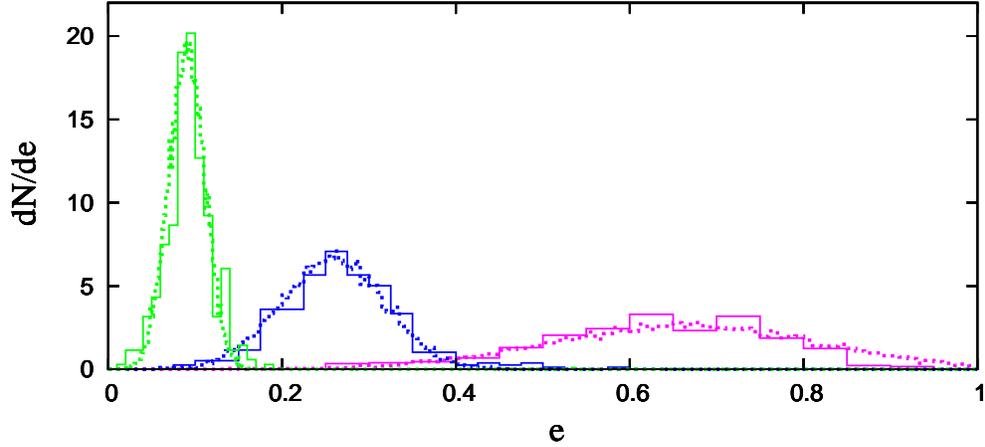} 
  \caption{Dynamical instability and asymptotic properties
of systems with two closely separated gas giant planets.  
All the models start with two planets with a total mass 
$(m_1+m_2)/M_*=0.006$ around a common host star.  Their 
initial orbits are nearly circular and coplanar 
with a unstable semimajor axis separation 
(ie $\Delta a < \Delta a_c$).  In the absence of gas drag, 
one planet in each system is eventually ejected while the 
other is retained with an changed orbit.  The eccentricity
distributions of the retained planets are plotted for 
three sets of models (with $\beta = m_2/m_1=0.1, 0.25$ 
and  0.5 from left to the right).  The solid and dotted 
lines represent results obtained from our N-body 
simulations and our analytic plus Monte Carlo prescription.}
  \label{fig:e_dis}
\end{figure}

In order to verify the validity of the analytic plus Monte Carlo
prescription, we compare the results generated with it and those 
obtained from a series of N-body simulations with a set of similar
initial conditions as those presented by \citet{FR08}. In all of 
these models, we consider the interaction between two planets 
with small initial eccentricities and inclinations.  We set their
initial semimajor axis to be $a_1 = 1$ and $a_2$ is randomly 
specified to be in the range of $[0.9 (1 + \Delta a_{\rm c}), 
(1 + \Delta a_{\rm c})]$ so that they generally undergo orbital 
crossings. 

We consider three series with $\beta = m_1/m_2=0.1, 0.25$ and  0.5.  
In all cases, $(m_1+m_2)/M_*= 0.006$. For each value of $\beta$, 
we carried out $\sim$300 N-body simulations with different 
random number seeds for initial relative orbital 
phases between the two planets. In the absence of gas drag, 
one planet is eventually ejected.  The differential 
eccentricity distribution function of the retained 
planet is represented by the solid lines in Figure 
\ref{fig:e_dis}. For the cases with $\beta = m_1/m_2
=0.1$ and $0.25$, the more massive planet 2 is generally 
retained. But, this mass preference is reduced 
for the $\beta=0.5$ cases.  

We also adopted the analytic plus Monte Carlo prescription 
to generated $10^4$ models for the same sets of $\beta$ values 
with the distribution in Eqs.~(\ref{eq:e_escjr20}) and 
(\ref{eq:ang_momentum20}). The results of the Monte Carlo 
calculations are represented by the dotted lines in Figure 
\ref{fig:e_dis}.  They are in good agreement with those of the 
N-body simulations.

Depending on the difference between their semimajor axes, 
two planet systems either rapidly undergo orbital crossings 
(if $\Delta a < \Delta a_c$) or completely avoid close
encounters. Some systems may evolve into unstable states 
through disk migration (which leads to a reduction in 
$\Delta a$) or gas accretion (which widens $\Delta a_c$).  In 
their natal disks, the eccentricities of gas giants are also
damped. In our population synthesis simulations, we include
the effect of eccentricity damping for the retained 
planet.  The efficiency of this effect depends on the
mass ratio between the planet and local disk
(see Appendix A1). 

Finally, the semimajor axis of the retained planet ($a_1$) is 
obtained from the conservation of energy, such that
\begin{equation}
\frac{m_1}{a_1} = \frac{m_1}{a_{1,0}} + \frac{m_2}{a_{2,0}},
\label{eq:energy_consv0}
\end{equation}
where $a_{j,0}$ is the semimajor axis before scattering. For 
computational simplicity, we neglect 1) the decrease in the 
semimajor axis change associated with the eccentricity damping 
of the retained planet, and 2) the finite kinetic energy of 
the ejected planet.

\subsubsection{Retention of both gas giants after close encounters 
between them}
\label{sec:tworetain}

If the trial eccentricities (generated at step 3) of both planets 
are less than unity, they would both be adopted as the asymptotic 
values.  The effect of the eccentricity damping due to the 
post-encounter disk-planet interactions is applied to both planets 
with the same prescription as that described in the previous section 
(i.e., in \S\ref{sec:twoeject}).  In \S 5.4, we also discuss the 
eccentricity damping due to disk gas accretion onto a planet.

Close encounters launch both planets into eccentric orbits
from nearly circular initial orbits through changes in 
their energy and semimajor axis.  We assume that a less 
massive body (represented by the subscript ``2'') is 
always scattered outward such that its semimajor axis prior 
to the close encounter $a_{2,0}$ become its new periastron
distance, such that 
\begin{equation}
a_{\rm 2} = \frac{a_{\rm 2,0}}{1 - e_2}.
\end{equation}
The new semimajor axis of the inwardly scattered planet ($a_1$) 
is given by the energy conservation such that
\begin{equation}
\frac{m_1}{a_1}=\frac{m_1}{a_{1,0}} + \frac{m_2}{a_{2,0}} 
- \frac{m_2}{a_{2}}.
\label{eq:energy_consv2}
\end{equation}

\subsection{Three gas giants case}
\label{sec:threegiants}
In contrast to the stability criterion in 
Eq.~(\ref{eq:deltaac}) for two planet systems, 
three planet systems may maintain relatively 
low-eccentricity states until gas in their natal
disk is severely depleted and then undergo close 
encounters after a time scale $\tau_{\rm cross}$.
The magnitude of $\tau_{\rm cross}$ is a sensitive 
function of their initial separation in semimajor 
axis $\Delta a$. 

In order to develop an easy-to-use and robust 
prescription for close encounters among three 
planets, we consider the situation that 
$\tau_{\rm cross}$ is smaller than the expected 
main sequence lifespan
of their host stars.  In these systems, eccentricities 
of the planets are always excited during the close 
encounters.  Since these events take place in the 
absence of residual gas drag, the planets' excited 
eccentricity can only be damped through 1) dissipative 
merger events, 2) tidal dissipation in the proximity
of their host stars, or 3) dynamical friction by a residual
population of planetesimals.  In this paper, we neglect the 
effect of the tidal dissipation and residual planetesimals.

Following a similar procedure from the previous section, 
we construct an analytic plus Monte Carlo prescription for 
the outcomes of close encounters and compare it with the 
result generated from N-body simulations of 
three-giant-planet systems which undergo orbital crossings 
in a gas-free 
environment \citep{Marzari02,Nagasawa08,Chatterjee08}. 
In this case, two or more planets ensure repeated 
encounters with a range of impact parameter. Physical 
collisions occur when the impact parameter is less
than the sum of the planets' radii.  Such merger events 
reduce the number of residual planets bound to the host 
star and consequently stabilize the remaining systems. 

Slightly wider scatterings, with impact parameter marginally
larger than the sum of planets' radii, lead to recoil velocities 
comparable to their surface escape speed. At a distance of 
several AU from their solar-type host stars, such strong 
orbital deflection generally leads to the escape of one 
planet from the gravitational potential of the host star.
The remaining two planets are retained with modest to 
high eccentricities and widely separated semimajor axes 
such that the residual systems are stabilized. 

We present below our analytic plus Monte Carlo prescriptions 
for the outcomes of close encounters in three-planet systems. 
Based on N-body simulations, we assume that this intense 
interaction leads to 1) mostly the ejection of one planet 
and 2) a significant fractional ($\sim 30\%$) incidents of 
physical collisions and merger events.  In both outcomes, 
the residual systems are left with two remaining planets. 
The main objective of our prescription is to evaluate the 
asymptotic eccentricities and semimajor axes of these 
two retained planets.

We justify the validity of our prescription based on the direct 
comparison of our results with those generated from the N-body 
simulations. The technical procedure essentially follows that 
constructed for the two planet case in \S\ref{sec:twogiants}, 
with some modifications applied to the inwardly and outwardly 
scattered retained planets. Their detailed description 
and physical justification are given in Appendix A2.  We outline 
the following basic steps.   
\begin{description}
\item[B1)] We first set initial conditions of the three giants 
and assign the identity of the most massive planet to be $j=1$, 

\item[B2-a)] We assign a probability ($30\%$) that two 
planets physically collide and merge with each other 
(assumed to always occur after each collision).
A merged planet is formed from a randomly selected 
pair of planets which are involved in the orbital crossings.
Its resultant semimajor axis and eccentricity are calculated 
under the assumed conservation of total mass, orbital energy 
and angular momentum. 

\item[B2-b)] We also assign a probability ($70\%$) that one of the
three planets is ejected while the other two residual planets 
are retained.  

For the ejection events,
\begin{description}
\item[B3)] 
we compute the maximum eccentricity  
$e_j ^{\rm max}$ for all three ($j=1, 2, 3$) planets, 
in accordance with the procedure described in step 2 
in Appendix A2. The magnitude $e_j ^{\rm max}$ of 
the most massive body tends to be
less than those of the other bodies.

\item[B4)] We select the planet with the largest 
value of of $e_j ^{\rm max}$ to be the ejector.

\item[B5)] We determine asymptotic eccentricities 
of the two retained planets to be
\begin{equation}
e_j  = 
\left\{
\begin{array}{ll}
\left(\frac{m_1 + m_{\rm ejc}}{m_1 + m_{j}}\right)^{1/2}
\left(\frac{R_1 + R_{\rm ejc}}{R_1 + R_{j}}\right)^{1/2} 
{\cal R}_j & ({\rm for} \;  j \ne 1), \\
\left(\frac{m_2}{m_1}\right)
\left(\frac{m_1 + m_{\rm ejc}}{m_1 + m_2}\right)^{1/2}
\left(\frac{R_1 + R_{\rm ejc}}{R_1 + R_2}\right)^{1/2} 
{\cal R}_j & ({\rm for} \; j = 1).
\end{array}
\right.
\label{eq:e_3body_ejc20}
\end{equation}
where $R_j$ is the physical radius of planet $j$ and
${\cal R}_j$ is a random number following a Rayleigh 
distribution of unit root mean square.  The above
expression is similar to that in Eq.~(\ref{eq:e_escj00})
with the consideration of incomplete stirring (see 
Appendix A2 for justification).

\item[B6)] Among the two retained planets, we select the 
inwardly scattered body with a probability distribution 
function which is weighted by the square of their masses.
\item[B7)] Based on the nature of close scattering, the 
new semimajor axis of the outwardly scattered body is 
prescribed to be
\begin{equation}
a_{\rm out}(1-e_{\rm out}) = \sqrt{a_{\rm max} a_{\rm min}} 
+ a_{\rm max}{\cal R}_j,
\label{eq:a_out0}
\end{equation}
where $e_{\rm out}$ is the excited eccentricity of the outer planet,
and $a_{\rm max}$ and $a_{\rm min}$ are the maximum and minimum
semimajor axes of the three planets in their initial state prior to 
orbital crossings.
\item[B8)] The semimajor axis of the inwardly scattered body is
determined by
\begin{equation}
\frac{m_{\rm in}}{a_{\rm in}} = E - \frac{m_{\rm out}}{a_{\rm out}},
\label{eq:a_in0}
\end{equation}
where $m_{\rm in}$ is the mass of the inner planet and $E$ is the total 
energy calculated from the initial semimajor axes of the three planets.
\end{description}
\end{description}

In order to compare the results of our prescription with those 
obtained from previous N-body simulations, we adopt a set of models
with three planets in nearly circular and coplanar orbits as 
initial conditions in step B1.  For these test cases, we consider 
the situation that orbital crossings have already been initiated after 
a timescale of $\sim \tau_{\rm cross}$ has elapsed, so that the 
initial orbital separations are arbitrarily set.
In the actual population synthesis simulations, $\tau_{\rm cross}$ 
is calculated and compared with the system time at each timestep 
(see Appendix A3).  

\begin{figure}[btp]
  \epsscale{1.1}       
 \plotone{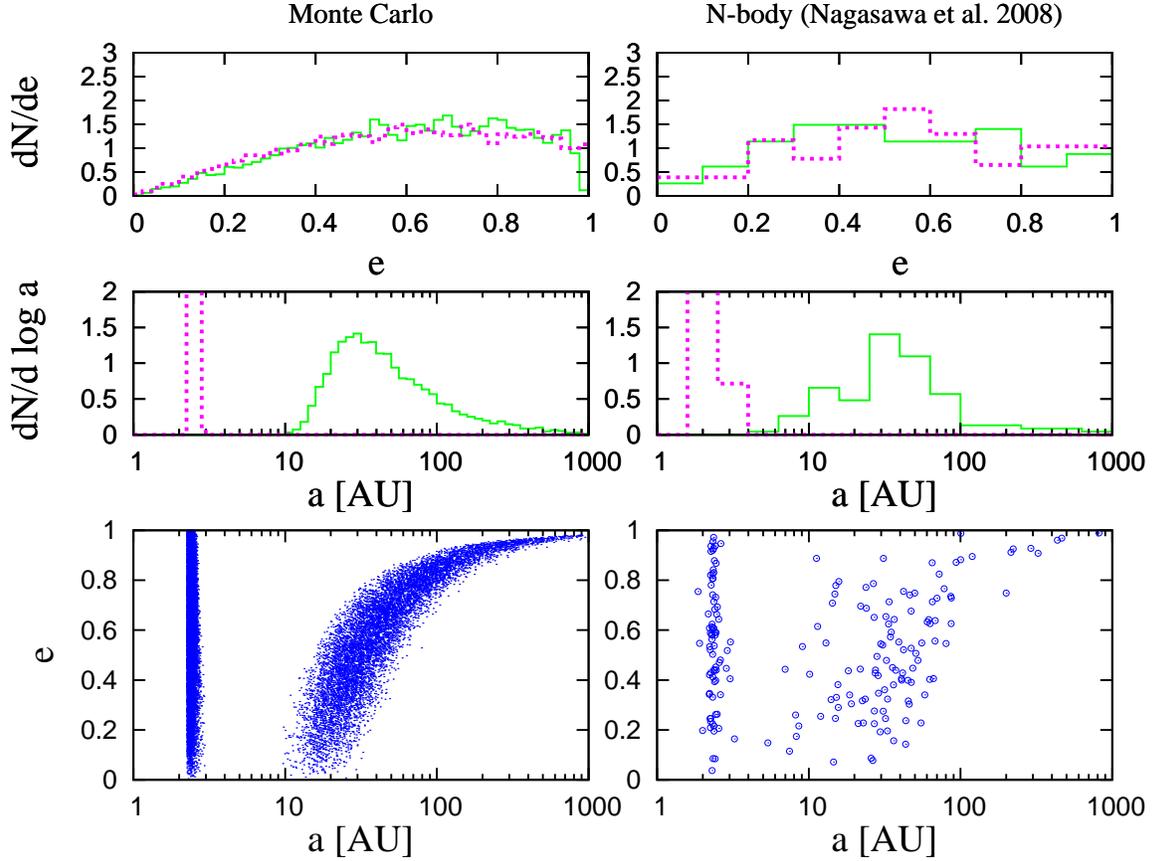} 
\caption{   
The eccentricity and semimajor axis distributions of 
surviving planets in the case of three Jupiter-mass 
planets with $a_1=5$AU, $a_2=7.25$AU, $a_3=9$AU and 
$e=0$ initially. The dotted and solid lines represent
the distributions of inner and outer planets in
the top and middle panels.
The left panels show the results of $10^4$ models generated with
our analytic plus Monte Carlo prescription and the right ones are
the results of 75 models obtained from \citet{Nagasawa08}'s 
N-body simulations.
}
  \label{fig:3giants}
\end{figure}

The comparison of our prescribed method and N-body simulations 
is shown in Figure \ref{fig:3giants} where we illustrate the 
eccentricity and semimajor axis distributions of the surviving planets 
for a test case with three Jupiter-mass planets ($m_1 = m_2 = m_3 
= M_{\rm J}$).  We adopt a set of initial conditions with $a_1=5$AU, 
$a_2=7.25$AU, $a_3=9$AU and $e=0$, in accordance with some existing
N-body simulations by \citet{Marzari02} and \citet{Nagasawa08} 
(their set N without tidal damping).

The left panels show the results of $10^4$ models generated with our 
analytic plus Monte Carlo prescription and the right panels illustrate
the results of 75 models of \citet{Nagasawa08}'s N-body simulations.
The initial orbital phase distributions of planets are modified 
in each of the N-body simulations. In the analytic plus Monte Carlo 
simulations, the seed for random number $({\cal R})$ generation
to determine eccentricity is changed in each simulation.

The N-body simulations are carried out in 3D, i.e.,
the orbital inclinations of each planet are also calculated.
In our analytic plus Monte Carlo simulations, although the
planets' inclinations are not explicitly computed, 
their distributions can be evaluated by $i$ (in radian) 
$\sim e/2$ \citep{Nagasawa08}.

In this equal-mass example, the top panel of Figure \ref{fig:3giants}
shows that the analytic plus Monte Carlo prescription produces 
a broad differential eccentricity distribution which is similar for 
the inner and outer planets.  We prescribed (in step B5) a Rayleigh's
distribution through $({\cal R}_j)$ which has the form $f(e)de = 
2 e \exp(-e^2)de$ (see Eq.~\ref{eq:e_3body_ejc20}).  The peak 
of this distribution function occurs at $1/\sqrt{2}\simeq 0.71$.

The middle panel of Figure \ref{fig:3giants} shows a bimodal 
semimajor axis distribution because inwardly scattered
planets are generally well separated from outwardly scattered 
planets. The middle panel of Figure \ref{fig:3giants} shows a 
correlation between semimajor axis ($a$) and eccentricity ($e$) 
of the outwardly scattered planets. This correlation arises 
because we assume (in step B7) that these planets are scattered 
from the region near the initial locations and their periastron 
radii are close to their original semimajor axes despite some 
outward diffusion before the scattering (see Appendix A2).
These results are all consistent with the results obtained by 
the previous N-body simulations.

\begin{figure}[btp]
  \epsscale{1.1}       
 \plotone{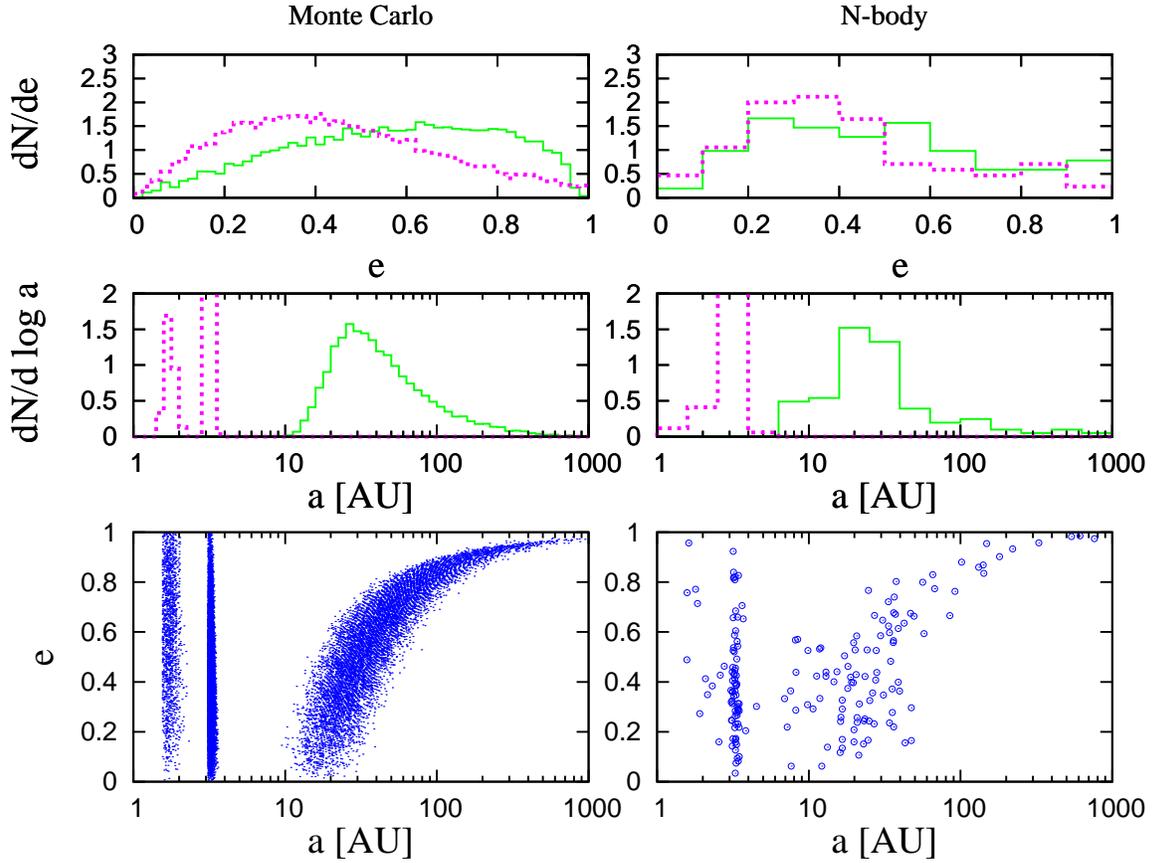} 
\caption{   
The eccentricity and semimajor axis distributions of 
the retained planets in the case of $m_1 = 2M_{\rm J}, 
m_2 = m_3 = M_{\rm J}$.  All other initial conditions 
are the same as those in Fig.~\ref{fig:3giants}.
}
  \label{fig:3giants_neql}
\end{figure}

\citet{Marzari02} also carried out simulations with $m_1 = 2 M_{\rm J}$ 
and $m_2 = m_3 = M_{\rm J}$.  They found that in 54 runs of 64 runs
(corresponds to $\sim 84\%$ probability), the more massive planet 
(with $m_1 = 2 M_{\rm J}$) is scattered inward and 
the inner retained planets have a broad eccentricity distribution peaked 
around $e \sim 0.3$.  Here, we carry out a similar simulation with the 
analytic plus Monte Carlo prescription to generate $10^4$ models.
These results show that the massive planet is scattered inward
in 8382 models with a peak eccentricity $1/2\sqrt{2} \simeq 0.36$ 
(in accordance with eq.~(\ref{eq:e_3body_ejc20})).  

For comparison purpose, we also simulated 100 N-body models.
For the massive planets' inward scattering probability
and their average peak eccentricities, 
the results obtained with the analytic-plus Monte Carlo 
prescription are almost consistent with those generated with the N-body simulation,
which provides support for the validity of the mass-weighted probability
function we have assumed. The middle panel in 
Fig.~\ref{fig:3giants_neql} show two sharp peaks in the semimajor 
axis distribution for the inwardly scattered planets.  These peaks are
also found in the N-body simulations. They are generated from
the requirement for energy conservation such that 
\begin{equation}
\frac{m_{\rm in}}{a_{\rm in}} \sim \frac{2M_{\rm J}}
{5{\rm AU}}+\frac{M_{\rm J}}{7.25{\rm AU}}+\frac{M_{\rm J}}{9{\rm AU}}.
\end{equation}
If the inner planet has a mass $m_{\rm in}=M_{\rm J}$ 
$a_{\rm in} \simeq 1.54$AU whereas $a_{\rm in} \simeq 3.08$AU
if $m_{\rm in}=2M_{\rm J}$.

\subsection{General case}

In the population synthesis simulations, multiple giant planets 
and rocky/icy planetary embryos co-exist.  Here we describe 
a summary of the prescriptions for dynamical interactions in 
the general case.  

During the early stages of their dynamical evolution, newly formed 
planets are embedded in and interact with their natal disks. In 
the presence of the disk gas, we only consider scattering 
between two planets when their orbital separation $\Delta a$ 
becomes less than the critical value $\Delta a_{\rm c}  \simeq 
2\sqrt{3} r_{\rm H}$ with the prescription given in \S2.2.

We calculate the planets' eccentricity damping timescale 
($\tau_{\rm damp}$), due to their gravitational interaction 
with the residual disk, and the orbit-crossing timescale 
($\tau_{\rm cross}$) for all pairs of existing planets 
in gas-free environment. As the disk gas is depleted, 
$\tau_{\rm damp}$ eventually becomes larger than 
$\tau_{\rm cross}$ and the planets' eccentricity grows 
due to secular excitation \citep{Iwasaki02}. It then 
becomes possible for planets with initial separation greater
than $\Delta a_{\rm c} \simeq 2\sqrt{3} r_{\rm H}$ to cross
each other's orbits.

Orbital crossings and close encounters among gas giants strongly 
affect the asymptotic global structure of planetary systems.
The gas giants' eccentricities are highly excited during their
orbital crossings.  The order of their semimajor axis is 
occasionally swapped.  Their secular resonances may sweep 
across the entire systems multiple times.  Most of the residual
small planets are either ejected or collide with their host
stars \citep{Matsumura13}.

For the evaluation on the outcome of planetary scattering after the 
severe disk-gas depletion, we apply the following prescription.
The details are presented in Appendix A3.
Note that the order of the prescription presented in this section is different
from those in Appendix A3, for the purpose of easier understanding.
\begin{description}
\item[C1)] We identify the "giant planets" from a list of all the planets 
in the system with the criteria that they have (i) a mass 
$m > 30 M_{\oplus}$ and (ii) $e_{\rm esc} > 1$ (Eq.~\ref{eq:e_escG_cond}).
\item[C2-a)] We apply the procedures for orbital crossings among the small 
planets (as outlined in Paper VI) if there is only one (or less) giant planet. 
\item[C2-b)] In systems with two (and only two) widely separated
(with $\Delta a > \Delta a_c$) giant planets, they retain their initial
kinematic properties (because they do not significantly perturb each other).  
Their dynamical influence on the less-massive planets
is computed independently, following the procedures in step C2.
\item[C2-c)] In systems with two (and only two) closely separated
(with $\Delta a < \Delta a_c$) giant planets, the outcome 
of their dynamical interaction is computed following steps A2-A3
in \S2.2.  We also remove all the small planets under the assumption
that they are ejected by the gas giants' sweeping secular resonances.
\item[C2-d)] In systems with 3 or more giant planets around the same host 
stars, we first evaluate $\tau_{\rm cross}$ for all pairs.
For this case,
\begin{description}
\item[C3-a)] if $\tau_{\rm cross}$ for all giant planet pairs is longer 
than the system time ($\tau_{\rm sys}$), we would neglect any interaction
between them and compute their influence on the low-mass planets 
following step C2-a.
\item[C3-b)] If $\tau_{\rm cross}$ for one or more pairs of giant
planets is less than $\tau_{\rm sys}$ but there are no triples
(overlapping pairs) with $\tau_{\rm cross} < \tau_{\rm sys}$,
we would determine the outcome of interaction between the giant
planets following step C2-b or C2-c.  
\item[C3-c)] Any groups of three giant planets around the same
host stars would be labelled as interacting triples if 
$\tau_{\rm cross} < \tau_{\rm sys}$ for at least two planet 
pairs.  In this case, we follow steps B2-B8 for the triple 
giants interaction case (see \S2.3).  All the small planets 
would be removed under the assumption that they are ejected 
by the triple giants' sweeping secular resonances.
\begin{description}
\item[C4)] After the kinematic properties of interacting
triple systems are modified to a set of new configuration
in the previous step, we re-calculate the $\tau_{\rm cross}$ 
for each pairs of giant planets.  Steps C2-C3 are repeated
until either 1) there are only two giant planets remaining 
around the host stars or 2) when $\tau_{\rm sys}+\tau_{\rm cross}
> 10^9$ yrs.
\end{description}
\end{description}
\item[C2-e)] For systems with more than 3 giant planets, we follow 
the same procedure as C3-a and C4 and eject all the small planets. 
But, we let only one giant planet be scattered inward and all other 
giant planets are assumed to be scattered outward.  
\end{description}

\section{Prescriptions of other physical processes in the population 
synthesis models}

We follow Paper VI for prescriptions of a) planetesimals' growth
through cohesive collisions, b) the evolution of planetesimal 
surface density,  c) embryos' type I migration, and their stoppage  
at the disk inner edge.  For the gas giants, d) the onset, rate, 
and termination (through gap opening or and global depletion)
of efficient gas accretion, and e) their type II migration, we 
follow prescriptions in Paper IV, except for a slight 
modification on gas accretion termination due to global depletion.
Here we present a brief summary.

\subsection{Disk models}
\label{sec:diskmodel}
We adopt the minimum mass solar nebula (MMSN) model \citep{Hayashi81} 
as a fiducial set of
initial conditions for planetesimal surface density ($\Sigma_d$) 
and introduce a multiplicative factor ($f_{d}$).
For gas surface density $\Sigma_g$, 
although we adopt the $r$-dependence of
steady accretion disk with constant $\alpha$ viscosity
($\Sigma_g \propto r^{-1}$),
rather than that of the original MMSN model ($\Sigma_g \propto r^{-1.5}$),
we scale $\Sigma_g$ by that of the MMSN at 10AU
with a scaling factor ($f_g$).
Following Paper IV, we set 
\begin{equation}
\left\{ \begin{array}{ll} 
\Sigma_d & = \Sigma_{d,10} \eta_{\rm ice} 
f_d (r/ {\rm 10 AU})^{-1.5}, 
\label{eq:sigma_dust} \\ 
\Sigma_g & = \Sigma_{g,10} f_{g} (r/ {\rm 10 AU})^{-1},
\label{eq:sigma_gas}
\end{array} \right.
\end{equation}
where normalization factors $\Sigma_{d,10} = 0.32 {\rm g/cm}^2$ and
$\Sigma_{g,10} = 75 {\rm g/cm}^2$ correspond to 1.4 times of
$\Sigma_g$ and $\Sigma_d$ at 10AU of the MMSN model, and the step
function $\eta_{\rm ice} = 1$ inside the ice line at $a_{\rm ice}$
(eq.~[\ref{eq:a_ice}]) and 4.2 for $r > a_{\rm ice}$ [the latter can
be slightly smaller ($\sim 2-3$) \citep{Pollack94}].
We specify an inner disk boundary where $\Sigma_g$ vanishes
and planetesimals' type I migration is arrested (see \S3.5).

Neglecting the detailed energy balance in the disk \citep{Chiang97, 
Garaud07}, we adopt the equilibrium temperature distribution of 
optically thin disks prescribed by \citet{Hayashi81} such that
\begin{equation}
T = 280 \left(\frac{r}{1{\rm AU}}\right)^{-1/2}
    \left(\frac{L_*}{L_{\odot}}\right)^{1/4} {\rm K},
\label{eq:temp_dist}
\end{equation}
where $L_*$ and $L_{\odot}$ are stellar and solar luminosity.
We set the ice line to be that determined
by an equilibrium temperature (eq.~[\ref{eq:temp_dist}]) in optically
thin disk regions,
\begin{equation}
a_{\rm ice} = 2.7 (L_\ast/L_\odot)^{1/2} {\rm AU}.
\label{eq:a_ice}
\end{equation}
The effects of evolution of the disk temperature and associated
migration of the ice line
adopting more realistic thermal disk structure
\citep{Garaud07, Oka11, Min11}
will be studied in a subsequent paper.

Dependence of disk metallicity is attributed to distribution of
$f_{d,0} = f_{g,0} 10^{{\rm [Fe/H]}_d}$, where $f_{d,0}$ and $f_{g,0}$
are initial values of $f_{d}$ and $f_{g}$, respectively. Due to
viscous diffusion and photoevaporation, $f_{g}$ decreases with time.
For simplicity, we adopt 
\begin{equation}
f_{g} = f_{g,0} \exp(-t/\tau_{\rm dep}),
\label{eq:gas_exp_decay}
\end{equation}
where $\tau_{\rm dep}$ is disk lifetime (for detailed discussion, see
Paper IV).  
The constant $\alpha$ self-similar solution obtained by
\citet{Lynden-Bell74} is expressed by
$\Sigma_g \propto r^{-1}$ with an asymptotic exponential
cut-off at radius $r_{\rm m}$ of the maximum viscous couple.
In the region at $r < r_{\rm m}$, $\Sigma_g$ decreases uniformly
independent of $r$ as the exponential decay does, although
the time dependence is different.
In the self-similar solution, $\Sigma_g$ at $r < r_{\rm m}$ 
decays as $\Sigma_g \propto (t/\tau_{\rm dep} + 1)^{-3/2}$.
In the exponential decay model that we adopt,
$\Sigma_g$ decays more rapidly as
$t/\tau_{\rm dep}$ becomes larger at $t > \tau_{\rm dep}$.
If the effect of photoevaporation is taken into account,
$\Sigma_g$ decays rapidly after it is significantly depleted, so that
the exponential decay partially mimics the effect of photoevaporation.
However, the final planet distributions hardly depend on
whether $\Sigma_g \propto (t/\tau_{\rm dep} + 1)^{-3/2}$ or 
$\Sigma_g \propto \exp(-t/\tau_{\rm dep})$.
In this paper, we adopt $\alpha = 10^{-3}$.
The parameter values of $f_{g,0}$, $\tau_{\rm dep}$, and [Fe/H]
are specified for each run.

\subsection{From Oligarchic Growth to Isolation}
\label{sec:oligar} 
On the basis of oligarchic growth model \citep{KI98,KI02}, growth rate
of embryos/cores at any location $a$ and time $t$ in the presence of
disk gas, is described by $dM_{\rm c}/dt = M_{\rm c}/\tau_{\rm c,acc}$
where, after correcting some typos in Paper IV, 
\begin{equation}
\tau_{\rm c,acc} = 3.5 \times 10^{5} \eta_{\rm ice}^{-1}
f_d^{-1} f_{\rm g}^{-2/5} 
\left( \frac{a}{1{\rm AU}} \right)^{5/2}
\left(\frac{M_{\rm c}}{M_{\oplus}} \right)^{1/3}
\left(\frac{M_\ast}{M_{\odot}} \right)^{-1/6}
{\rm yrs},
\label{eq:m_grow0}
\end{equation}
where $M_{\rm c}$ is the mass of the embryo (core), and we set the mass of the typical
field planetesimals to be $m=10^{20}$g.  

If type I migration is not effective, embryos would formed through 
oligarchic growth and attain a local isolation mass, $M_{\rm c,iso}
= 2 \pi r \Sigma_d \Delta a_{\rm c}$ where the full width of feeding 
zone of an embryo with a mass $M_{\rm c}$ is given by \citep{KI98,KI02}
\begin{equation}
\Delta a_{\rm c} \simeq 10 r_{\rm H} = 
10 \left( \frac{2M_{\rm c}}{3M_{\ast}} \right)^{1/3}a, 
\label{eq:m_iso}
\end{equation}
and $r_{\rm H}$ is the Hill radius for two bodies with comparable
masses.  From equations (\ref{eq:sigma_dust}) and (\ref{eq:m_iso})
we find 
\begin{equation}
M_{\rm c,iso} \simeq
0.16 \eta_{\rm ice}^{3/2} f_d^{3/2} 
\left(\frac{a}{1\mbox{AU}}\right)^{3/4} 
\left(\frac{M_\ast}{M_{\odot}} \right)^{-1/2} M_{\oplus}.
\label{eq:m_iso0}\end{equation}

However, type I migration may induce embryos to migrate 
before they acquire all the residual planetesimals within
their feeding zone and acquire their isolation mass.
Type I migration may also be stalled near some 
trapping radius.  At these locations, the congregation of 
planetesimals may increase the magnitude of $\Sigma_d$ and 
enlarge the magnitude of $M_{\rm c,iso}$.

We compute the evolution of $\Sigma_d$ distribution due to 
accretion by all the emerging embryos in a self-consistent 
manner. The growth and migration of many planets are integrated 
simultaneously with the evolution of the $\Sigma_d$-distribution. 
We set up linear grids for $f_d$ across the disk with typical 
width of $\sim 10^{-3}$AU. We introduce a population of seed 
embryos, all with an initial mass $10^{20}$g (i.e., that
of the residual planetesimals) and compute their mass accretion
rate.  In the inner disk region, we set their initial separation 
of the full feeding-zone width ($ \Delta a_{\rm c} =10 
r_{\rm H}$) of embryos with local asymptotic isolation mass 
$M_{\rm c, iso}$.  

The planetesimals' growth time scale increases rapidly with 
their distance from the central stars and embryos in the outer 
disk region of the disk are unlikely to attain a local isolation 
masses within the life span of their host stars.  There, we 
place seed embryos which are separated by feeding zone width
of embryos with masses evaluated (using Eq \ref{eq:m_grow0})
for the local $\Sigma_d$ after $t \sim 1$ Gyr.  We follow the 
growth of the seed embryos due to planetesimal accretion
in accordance with Eqs.~(\ref{eq:m_grow0}) and (\ref{eq:m_iso}). 
The planetesimals' mass accreted by the embryos is uniformly 
subtracted within the embryos' current feeding zone.  We follow 
the evolution of $\Sigma_d$ throughout the disk and use its
values to evaluate embryos' local accretion rates.  We also use
$\Sigma_d (r, t)$ to estimate the strength of planetesimals' 
dynamical friction on the embryos. 

During the early phase of evolution, embryos are embedded in their
natal disks.  Despite their mutual gravitational perturbation, the
embryos preserve their circular orbits due to the gravitational drag 
from disk gas\citep{Artymowicz93, Ward93} and dynamical friction 
from the residual planetesimals\citep[e.g.,][]{Stewart00}. However, 
after the disk gas is severely depleted, the efficiency of eccentricity
damping mechanism is reduced.  The embryos' eccentricity grow until
they cross each others' orbits.  We use the prescription in Paper VI
and Appendix A3 to compute the occurrence and consequence of giant 
impacts between embryos.

\subsection{Type I migration}
\label{sec:migrat}

Type I migration of an embryo
is caused by the sum of tidal torque 
from the disk regions both interior and exterior to the embryos.  
The rate and direction
of embryos' migration are determined by the differential Lindblad 
and corotation torques. 
While  a conventional formula of type I migration 
assuming locally isothermal disks \citep[e.g.,][]{Tanaka02} shows that
the migration is always inward, recent developments
of type I migration of isolated embryos in non-isothermal disks 
\citep[e.g.,][]{Paardekooper11} show 
the magnitude and sign of the tidal torque, 
especially that due to corotation resonances is a sensitive function 
of the surface density and temperature distribution of the disk gas. 

Since the properties of migration depend on detailed 
thermal/dynamical structure of disks and "saturation" degree
of the corotation torque
\citep[e.g.,][]{Kretke12}, substantial discussions are required for
the effects of type I migration in non-isothermal disks.
The pace of individual embryos' type I migration in dense multiple 
planet systems remains uncertain because mutual perturbation between 
nearest neighbors may modify their Lindblad and especially corotation torque.  
On the other hand, as shown below, the eccentricity distributions
that we are primarily interested in here
are not sensitive to a change in the formula of type I migration.
Thereby, in order to highlight the validity of our treatment of dynamical
interaction between multiple planets in this paper, we 
postpone the detailed discussions on the non-isothermal
type I migration to a subsequent paper.
While we also show a result with a non-isothermal formula,
in most of the results we present here, 
we use a conventional formula 
of type I migration in isothermal disks
derived by \citet{Tanaka02}
with a scaling factor $C_1$: 
\begin{equation} 
\begin{array}{ll}
\tau_{\rm mig1} & 
{\displaystyle
= \frac{a}{\dot{a}} 
= \frac{1}{C_1}
  \frac{1}{3.81}
  \left(\frac{c_s}{a \Omega_{\rm K}}\right)^{2} 
  \frac{M_*}{M_p}
  \frac{M_*}{a^2 \Sigma_g}
  \Omega_{\rm K}^{-1} }\\
 & 
{\displaystyle
  \simeq 1.5 \times 10^5 \times \frac{1}{C_1 f_g} 
  \left(\frac{M_{\rm c}}{M_{\oplus}} \right)^{-1} 
  \left(\frac{a}{1{\rm AU}}\right) 
  \left(\frac{M_*}{M_{\odot}}\right)^{3/2}
  \;{\rm yrs}. } 
\end{array}
\label{eq:tau_mig1} 
\end{equation} 
The expression of \citet{Tanaka02} corresponds to $C_1 = 1$, and
for slower migration, $C_1 < 1$. 
We study a dependence of the eccentricity distribution
by changing a value of $C_1$, 
rather than using variations of non-isothermal formula.

We assume type I migration 
ceases inside the inner boundary of the disk, because
$f_g$ is locally zero there.  For computational 
convenience, we set the disk inner boundary to be the edge of the 
magnetospheric cavity at $\sim 0.04$ AU.

\subsection{Formation of gas giant planets}

Prescriptions for formation of gas giant planets are the same as those
used in Paper IV, except for slight modification for reduction and 
termination of gas infall. Embryos are surrounded by gaseous envelopes
when their surface escape velocity becomes larger than the sound speed 
of the surrounding disk gas.  When their mass grow (through planetesimal 
bombardment) above a critical mass 
\begin{equation}
M_{\rm c,hydro} \simeq
10 \left( \frac{\dot{M}_{\rm c}}{10^{-6}M_{\oplus}/ {\rm yr}}\right)^{0.25}
M_{\oplus},
\label{eq:crit_core_mass}
\end{equation}
both the radiative and convective transport of heat become sufficiently 
efficient to allow their envelope to contract dynamically\citep{Ikoma00}.  

In the above equation, we neglected the dependence on the opacity in the
envelope (see Paper I and \citet{Hori10}).  In regions where the 
cores have already acquired isolation mass,  their planetesimal-accretion 
rate $\dot M_{\rm c}$ would be much diminished \citep{Ikoma00, Zhou07} 
and $M_{\rm c,hydro}$ can be
comparable to an Earth mass $M_\oplus$.  But, gas accretion also
releases energy and its rate is still regulated by the efficiency of
radiative transfer in the envelope such that
\begin{equation}
\frac{dM_p}{dt} \simeq \frac{M_p}{\tau_{\rm KH}},
\label{eq:mgsdot}
\end{equation}
where $M_{\rm p}$ is the planet mass including gas envelope.  

In Paper I, we approximated the Kelvin-Helmholtz contraction timescale
$\tau_{\rm KH}$ of the envelope with
\begin{equation}
\tau_{\rm KH} \simeq \tau_{\rm KH1}
\left(\frac{M_p}{M_{\oplus}}\right)^{-k2},
\label{eq:tau_KH}
\end{equation}
where $\tau_{\rm KH1}$ is the contraction timescale for $M_p=M_{\oplus}$. 
Since there are uncertainties associated with dust sedimentation and 
opacity in the envelope\citep{Pollack96, Helled08, Hori11}, we adopt a range 
of values $\tau_{\rm KH1} = 10^8-10^{10}$ years and $k2 = 3$--4.
Here we fix $k2 = 3$ and change $\tau_{\rm KH1}$ (In a nominal case, 
we use $\tau_{\rm KH1} = 10^9$ years).

Equation~(\ref{eq:mgsdot}) shows that $dM_{\rm p}/dt$ rapidly increases 
as $M_p$ grows.  But, it is limited by the global gas accretion rate 
throughout the disk and by the process of gap formation near the 
protoplanets' orbits.  The disk accretion rate is
\begin{equation}
\dot{M}_{\rm disk} \simeq 3 \pi \Sigma_g \nu
\simeq 3 \times 10^{-9} f_g \left(\frac{\alpha}{10^{-3}}\right)  
\; [M_{\odot}/{\rm yr}].
\end{equation}
During the advanced stage of disk evolution, we assume both 
$\dot{M}_{\rm disk}$ and $\Sigma_g$ evolves $\propto 
\exp(-t/\tau_{\rm dep})$ where $\tau_{\rm dep}$ is the gas 
depletion time scale.  The rate of accretion onto the cores
cannot exceed ${\dot M_{\rm disk}}$.

An (at least partial) gap is formed when the planets' tidal torque
exceeds the disk's intrinsic viscous stress \citep{LP85}.
This viscous condition for gap formation is satisfied for planets with
\begin{equation} 
M_p > M_{\rm g,vis} 
\simeq 30 \left(\frac{\alpha}{10^{-3}}\right) 
\left(\frac{a}{1{\rm AU}}\right)^{1/2} 
\left(\frac{L_\ast}{L_{\odot}}\right)^{1/4} M_{\oplus}.
\label{eq:m_gas_vis} 
\end{equation} 
Then, type I migration is switched to type II migration.
Fluid dynamical simulations
\citep[e.g.,][]{D'Angelo03,Lubow08} show that
some faction of gas still flows into the gap.
According to this,
we allow a protoplanet to continue accreting the residual gas 
which flows past it, as shown below.
These fluid dynamical simulations, however, also show that
the accretion rate rapidly decreases with $M_p$
after $M_p$ exceeds a Jupiter mass.
According to this result and the analysis by \citet{Dobbs-Dixon07},
we completely terminate gas accretion, 
when the planet's Hill radius ($r_{\rm H}$) 
becomes larger than 2 times of disk scale height ($H$),
which corresponds to (extended) thermal condition 
\citep{LP85}, that is,
\begin{equation}
M_p > M_{\rm g,th}
\simeq 0.95 \times 10^3 \left(\frac{a}{1{\rm AU}}\right)^{3/4}
\left(\frac{L_\ast}{L_\odot}\right)^{3/8}
\left(\frac{M_\ast}{M_\odot}\right)^{-1/2} M_\oplus.
\label{eq:m_gas_th}
\end{equation}

In general, our prescriptions for gas accretion rates onto the cores are 
\begin{equation}
\frac{dM_p}{dt} = f_{\rm gap} \dot{M}_{p, {\rm no gap}},
\end{equation}
where $\dot{M}_{p, {\rm no gap}}$ is that in the absence of any feedback on
the disk structure, i.e., that without the effect of gap opening,
\begin{equation}
\dot{M}_{p,{\rm no gap}} = \min \left(\frac{M_{\rm p}}{\tau_{\rm KH}}, 
\dot{M}_{\rm disk} \right),  
\end{equation}
and $f_{\rm gap}$ is a reduction factor due to gap opening,
\begin{equation}
f_{\rm gap} = 
\left\{
\begin{array}{ll}
1 & [{\rm for} \; M_ p <  M_{\rm g,vis}] \\
\displaystyle{
\frac{\log M_p - \log M_{\rm g,vis}}{\log M_{\rm g,th} - \log M_{\rm g,vis}} 
} &
[{\rm for} \; M_{\rm g,vis} < M_ p <  M_{\rm g,th}] \\
0 & [{\rm for} \; M_ p >  M_{\rm g,th}].
\end{array}
\right.
\end{equation}
The formula for $M_{\rm g,vis} < M_ p <  M_{\rm g,th}$ is constructed 
to avoid any abrupt truncation. Some gas may leak through the gap.  But,
for planets with $M_p > M_{\rm g,th}$\citep{D'Angelo03}, the diffuse 
gas flows along horseshoe streamlines without being accreted by the 
planet \citep{Dobbs-Dixon07}.  Further discussion on this issue will be
discussed in a future paper.

In Papers I-III, we incorporated the effect of the global depletion 
by setting a limiting value for the planet's mass to be 
$M_p < M_{\rm g,no iso} \sim \pi a^2 \Sigma_g$.  This prescription 
was modified in Paper IV, in which we constrained $M_p < \int_0^{2a} 
2 \pi a \Sigma_g da$.   These previous prescriptions do not take into
account of the viscous diffusion of gas from other disk regions.
In order to consider the possibility that gas accretion may also
occur in the inner disk regions where the local gas content is limited, 
we evaluate in this paper both $\dot{M}_{p,{\rm no gap}}$ and 
$\dot{M}_{\rm disk}$ using the instantaneous values of $\Sigma_g$ $(\propto 
\exp(-t/\tau_{\rm dep}))$. The quenching of gas accretion due to the 
disk's global depletion is taken into account without any additional 
specification.  This modification enables us to compute the gas accretion
rate onto cores with relatively small semimajor axis.

\subsection{Type II migration}

During the gap formation, embedded gas giants adjust their
positions in the gap to establish a quasi equilibrium between 
the torque applied on them from the regions of the disk both 
interior and exterior to their orbits.  Subsequently, as the 
disk gas undergoes viscous diffusion, this interaction leads
to type II migration.  

We assume that planets undergo type II migration after they have 
accreted a sufficient mass to satisfy the viscous 
($M_{\rm g,vis} < M_ p$) condition for gap formation,
as commented in section 3.4.

While $M_p$ increases, the disk mass declines due to stellar
and planetary accretion and photoevaporation.  While the disk
mass exceeds $M_p$ (during the disk-dominated region), planets' 
type II migration is locked with the viscous diffusion of the 
disk gas.  During the advanced stages of the disk evolution 
when its mass become smaller than $M_p$ (during the 
planet-dominated region), the embedded planets carry a major share
of the total angular momentum content.  A significant fraction 
of the total (viscous plus advective) angular momentum flux 
transported by the disk gas is absorbed by the planet in its 
orbital evolution (for detailed discussion, see Paper IV). 

For the disk-dominated regime,
the migration timescale is given by
\begin{equation}
\begin{array}{ll}
\tau_{\rm mig2,disk} 
 & {\displaystyle  = \mid \frac{a}{\dot{a}} \mid 
= \frac{a}{(3/2) \nu / a} }\\
 & {\displaystyle \simeq 0.7 \times 10^5
\left(\frac{\alpha}{10^{-3}}\right)^{-1}
\left(\frac{a}{1{\rm AU}}\right)
\left(\frac{M_*}{M_{\odot}}\right)^{-1/2}}
\;{\rm yrs}.
\end{array}
\label{eq:tau_mig2a} 
\end{equation}
For the temperature distribution given by Eq.~(\ref{eq:temp_dist}), 
$\nu \propto r$ and $\tau_{\rm mig2,disk}$ reduces to 
\begin{equation}
\tau_{\rm mig2,disk} = \frac{a}{r_{\rm m}} \frac{2r_{\rm m}^2}{3\nu(r_{\rm m})}
\simeq \frac{a}{r_{\rm m}} \tau_{\rm dep},
\label{eq:tau_mig2a2}
\end{equation}
where $\tau_{\rm dep}$ is global disk viscous-evolution time scale 
and $r_{\rm m}$ is the disk radius where the viscous couple attains 
a maximum value.  (The magnitude of $r_{\rm m}$ is approximately 
the characteristic disk size). In general, $\tau_{\rm mig2,disk}$ 
is a fraction of the disk lifetime except for the planets which 
formed near $r_{\rm m}$.

Observationally inferred disk lifetime is 1-10Myrs.  If
this lifetime is comparable to the disk viscous-evolution
time scale, $r_{\rm m}\sim (15-150) \times (\alpha/10^{-3})$AU. 
In this limit, Eq.~(\ref{eq:tau_mig2a2}) would imply
that many gas giants form just beyond the snow line migrate to
the proximity of their host stars and become hot Jupiters
unless they emerged during the advanced (planet-dominated) 
stage of disk evolution or the type II tidal torque is 
significantly reduced by the disk flow through the gap
or by the mutual interaction between multiple planets and 
their natal disks.  

In Paper IV, we showed that the type II migration time scale
in the planet-dominated regime can be evaluated in terms of 
\begin{equation}
\begin{array}{ll}
\tau_{\rm mig2,pl} 
 & {\displaystyle = \frac{1}{C_2} 
\frac{(1/2)M_p a^2 \Omega_{\rm K}(a)}{\dot{M}_{\rm disk} 
r_m^2 \Omega_{\rm K}(r_m)}  = \frac{1}{2C_2}  
\left(\frac{a}{r_{\rm m}}\right)^{1/2} 
\frac{M_p}{\dot{M}_{\rm disk}}}
\\
 & {\displaystyle \simeq 5 \times 10^5 f_g^{-1} 
\left(\frac{C_2 \alpha}{10^{-4}} \right)^{-1} 
\left(\frac{M_{\rm p}}{M_{\rm J}} \right) 
\left(\frac{a}{1{\rm AU}}\right)^{1/2} 
\left(\frac{r_{\rm m}}{10{\rm AU}}\right)^{-1/2} 
\left(\frac{M_*}{M_{\odot}}\right)^{-1/2}}
\;{\rm yrs}, 
\end{array}
\label{eq:tau_mig2b} 
\end{equation} 
where $C_2$ is an efficiency factor associated with 
the degree of asymmetry in the torques between the inner 
and outer disk regions.  If the inner disk is severely 
depleted, $C_2 = 1$.  But, in evolving disks with comparable
surface density on either sides of multiple planets, the 
degree of torque asymmetry has large uncertainties in both
the disk and planet dominated regimes.  Although the 
reduction factors for the disk and planet-dominated regimes
may be independent from each other, we adopt the same 
reduction factor $C_2$ for both regime and treat the factor 
$C_2$ as a model parameter.  In the results shown in this 
paper, we set $C_2=0.1$.  The kinematic distribution of the
planets are weakly affected by the choice of $C_2$, as long 
as $C_1 \ga 0.03$.  The eccentricity distribution of the
emerging gas giant planets is not strongly affected by the 
magnitude of $C_2$.

\subsection{Resonant capture}
\label{sec:resona}

The pace of both type I and II depends on both the disk structure
and $M_p$. As different-mass planets migrate, the separation 
between their semimajor axis may either increase or decrease
through divergent and convergent migration respectively.
Multiple planets may also converge near some special disk 
locations such as the edge of the magnetospheric cavity 
near the inner disk boundary.  As the migration of first-born 
planets is stalled at these trapping radii, later-generation 
planets may continue to arrive.

The planets gradually converging orbits capture each 
other on their mutual mean-motion resonances provided
their migration time scale across the resonances' width 
is shorter than the resonances' libration time scales.
After they enter into their mean motion resonances, these 
planets have a tendency to migrate together while maintaining 
the ratio of their semimajor axes \citep{McNeil,Ogihara09}.
In their simulations, \citet{Ogihara09} and \citet{Ogihara10} 
found that outside the disk inner boundary at $\sim 0.1-0.3$AU,
it is possible for a few dozen rocky planetary embryos to 
be trapped in adjacent embryos' mean-motion resonances.
During and after the depletion of the disk gas, their 
orbits become unstable, they undergo orbit crossing and 
cohesive collisions to merge into a few super-Earths.

In Paper VI, we implemented a prescription for resonant capture
between rocky/icy planets into our sequential planet formation model.
With this prescription, we constructed population synthesis models
for multiple super-Earth systems. These simulations showed 
a rich population of embryos accumulate near the inner disk 
boundary.  Although differential migration leads to embryos' 
orbital convergence, stochastic secular perturbation between
multiple embryos induces orbital diffusion.  Instead of 
locking into the most widely separated 2:1 mean motion 
resonances, the nearest neighbors are packed into low-order 
main-motion resonance with semimajor axis separations 
comparable to several $r_{\rm H}$. The Hill radius
$r_{\rm H}$ is defined to be 
$r_{\rm H} = ((m_i + m_j)/3M_{\ast})^{1/3} a$ where
$m_i$ and $m_j$ are the masses of the closest pairs, 
$M_*$ is the host star's mass, and $a=\sqrt{a_i a_j}$
is mean semimajor axis of the pair.

For computational simplicity, we adopted in Paper VI, 
a prescription that, as a consequence of converging
migration, the closest embedded embryo pairs enter 
into and are locked in mean motion resonances with 
a separation of $5r_{\rm H}$ between their semimajor axes.
After capture each other, the resonant embryos migrate
in steps while maintaining their semimajor axis ratios 
and sharing angular momentum loss associated with the
gravitational interactions between the disk and individual 
embryos.    

In the present paper, we consider the possibility
that the converging differential migration may proceed
so rapidly that the neighboring embryos may not have
time to capture each other into mean motion resonances.
We evaluate the condition for a dynamical equilibrium 
in which convergence time scale (due to differential 
migration) matches with the mean motion libration time scale
(see Paper VI).  If this equilibrium is established with  
an orbital separation which is less than $\Delta a_{\rm c}  
\simeq 2\sqrt{3} r_{\rm H}$, we assume the converging embryos
will collide and coalesce (see \S2.2).  

We further extend the resonant capture condition for planet pairs 
including both super-Earths and gas giants.  Based on the rate of
orbital decay due to gas drag, \citet{Shiraishi08} derived a 
criterion for non-resonant planetesimals to enter the feeding 
zone of a growing gas giant.  Here we adopt their criterion for 
the embryos, the migration of which is determined by their 
tidal interaction with the disk gas.  We replace the
migration rate due to gas drag 
with differential speed between type II migration 
of a giant and type I migration of rocky/icy embryos.

In the case that a planet (an embryo or another giant) enters 
a feeding zone of a gas giant, we apply the prescription in 
\S2.1 to determine the outcome of their encounters.  Although 
the treatment in \S2.1 is constructed for scattering between 
two gas giants, the same prescriptions are also applicable 
for the case where an embryo is scattered by a giant.  In this 
limit, the perturbations from the giant is so strong and dominant
that these events tend to result in the ejection of the embryos 
or in widely separated orbits rather than cohesive collisions.

In Paper VI, we assumed that the disk inner edge is a rigid 
wall for type I migration of embryos.  This assumption is based 
on a strong "eccentricity trapping" effect\citep{Ogihara10}.
The torque exerted onto the innermost embryo in an eccentric
orbit is so strong that it can halt the type I migration and
offset the angular momentum loss of several outer embryos.
The magnitude of torque associated with embryos-planet interaction
depends on the planet's mass.  The condition for a single planet 
(with a mass $m_1$) to frustrate the inner migration of other
embryos (with masses $m_j$ $(j \ge 2)$) at the inner 
boundary \citep{Ogihara10} is
\begin{equation}
\sum_{j=1} m_j^2 < \frac{5}{C_1} m_1^2.
\label{eq:ecc_trap}
\end{equation}
Leakage of innermost embryo across the disk inner boundary 
would be imposed if the above condition is not satisfied. 
We also allow the leakage of embryos across the disk inner
boundary if they lie in the path of an inwardly migrating 
gas giant planet.

\section{Simulated individual systems}

Before we enter into an comprehensive discussion on the 
statistical properties and population synthesis of 
multiple-planet systems, we first highlight the dynamical
interaction among several bodies around common host stars 
with a few sample simulations of individual systems.  
Such studies have not been possible with the 
"one-planet-per-disk-model'' prescription in which planet-planet
interactions are neglected (e.g., Papers I-V and the models
of \citet{Mordasini09,Mordasini09b}).

\begin{figure}[btp]
\epsscale{1.0}       
\plotone{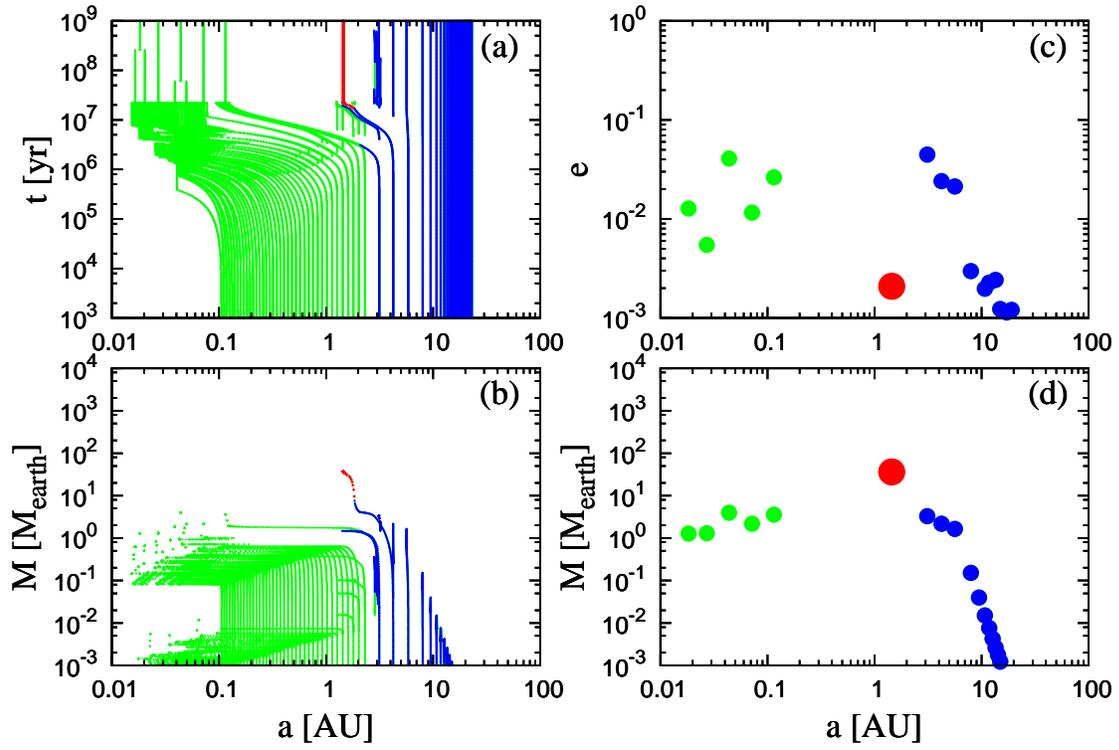} 
\caption{   
Growth and migration of planets in a system
with $C_1 = 0.1$ and $f_{d,0}=2.0$.
In the color version, the green, blue and red lines represent 
planets with their main component is rock, ice and gas, respectively.
}
  \label{fig:obt}
\end{figure}

For illustration purpose, we choose for Model 1, 
$C_1=0.1$, $f_{g,0} = f_{d,0} = 2.0$ ([Fe/H]$=0$),
$\tau_{\rm KH1}=10^9$ years, $\tau_{\rm dep} = 3 
\times 10^6$ years around a solar-mass star ($M_* = 1 M_{\odot}$).
Figure~\ref{fig:obt}a and b show time and mass evolution of 
planets for Model 1.  The green, blue and red lines represent 
rocky, icy, and gas giant planets with their main component being
rock, ice and gas, respectively.  Although rocky and icy
planets are formed respectively interior and exterior to the 
snow line, their migration and dynamical interaction 
may lead to their spatial diffusion.  The main composition 
of the planets may change through gas or planetesimal accretion 
or embryo-embryo collisions (giant impacts). 

The results in Panel b show that at 0.1--1AU small embryos 
grow {\it in situ} until their masses reach $\sim 0.1M_{\oplus}$
--$1M_{\oplus}$ and then undergo type I migration to 
accumulate near the inner boundary of their natal disks.  
We can determine the critical embryo mass for transition 
from local mass growth to efficient type I migration by 
matching Eqs.~(\ref{eq:m_grow0}) and (\ref{eq:tau_mig1}) 
such that
\begin{equation}
M_{\rm c,max} \simeq 1.0 
\left(\frac{C_1}{0.1}\right)^{-3/4}
\left(\frac{f_{g}}{2}\right)^{3/10}
\left(\frac{\eta_{\rm ice} f_{d}}{f_{g}}\right)^{3/4}
\left( \frac{a}{1{\rm AU}} \right)^{-9/8}
\left(\frac{M_\ast}{M_{\odot}} \right)^{5/4} M_{\oplus}.
\label{eq:m_c_max}
\end{equation}
A few small ($10^{-3}M_{\oplus}$--$10^{-2}M_{\oplus}$) embryos 
are also induced to undergo orbital decay.  The orbital evolution
of these low-mass embryos is induced by the perturbation from 
the migrating planets.  

Many resonant embryos accumulate in the vicinity of the 
disk inner boundary which is set to be 0.04 AU (Panel a,
Figure ~\ref{fig:obt}).  These planets are preserved in 
the disk region provided the condition (\ref{eq:ecc_trap}) 
is satisfied.  But when the total mass of the trapped 
planets in the proximity of the inner disk edge exceeds 
the critical mass given by eq.~(\ref{eq:ecc_trap}), innermost 
planets are driven to cross the disk inner edge until the
retention condition in Eq .~(\ref{eq:ecc_trap}) is restored.
In this model, more than half of the embryos arrived in 
the inner disk region eventually cross its inner boundary.
We assume that after planets migrate inside the 
central magnetospheric cavity in the disk, their tidal
and magnetic interaction with their host stars provide 
an effective torque to induce them to undergo further 
orbital decay.  Note that in the weak-field limit, the 
size of the magnetospheric cavity is much smaller and 
the disk inner boundary is adjacent to the stellar surface.  
In this limit the innermost planets are unlikely to survive 
(Paper VI).  

In the intermediate disk region, a migrating core attains 
$M_p \sim 5 M_{\oplus}$ just outside the ice line at 
$a \sim 3$AU.  This core starts its runaway gas accretion without
any significant type I migration.  When it evolves into a gas 
giant with a surrounding gap, it undergoes type II migration which
is much slower than type I migration.  The emerging gas giant 
scatter and eject nearby embryos (they are represented by 
the lines which are discontinued at $t \sim 2 \times 10^7$ yrs.)

\begin{figure}[btp]
\epsscale{1.0}       
\plotone{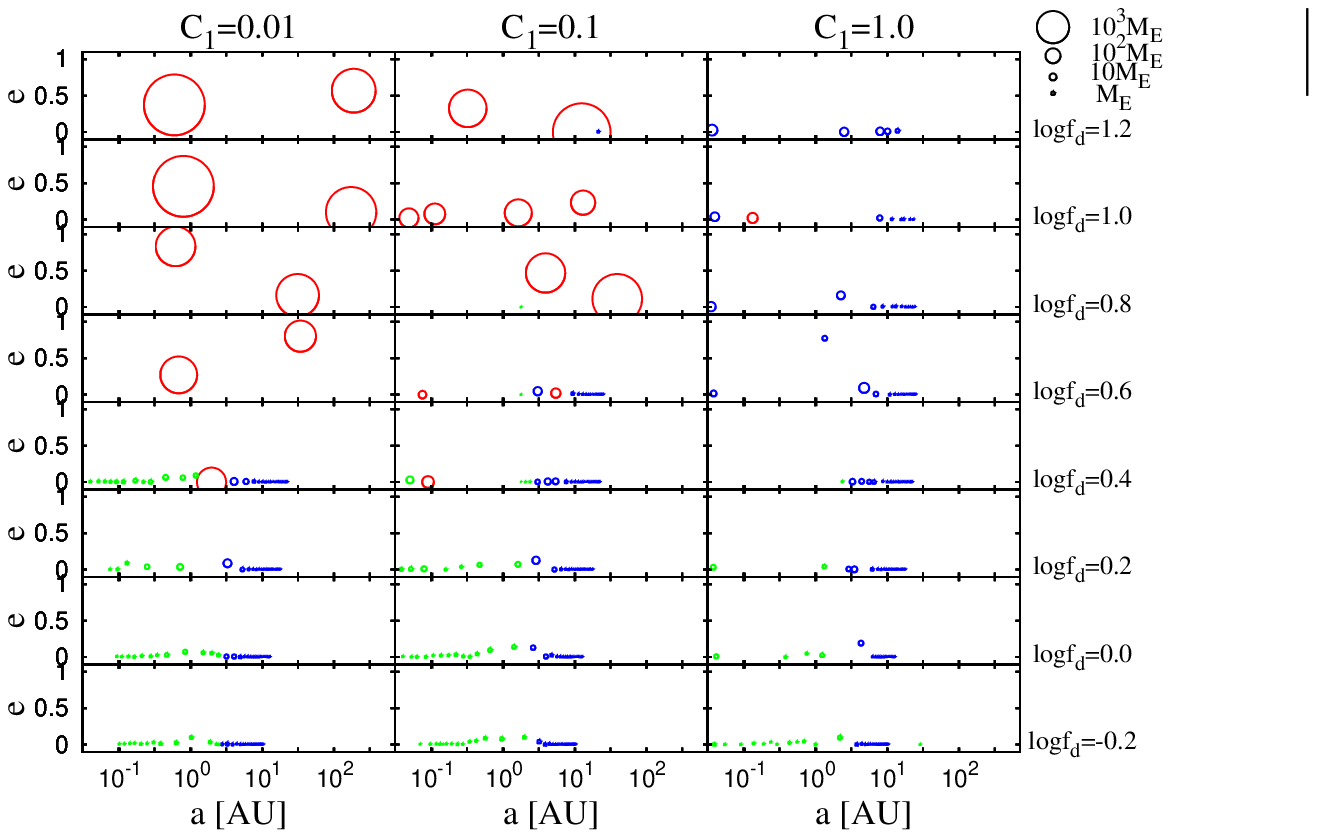} 
\caption{   
Planetary systems as functions of $C_1$ and $f_{d,0}$.
The radii of circles are proportional to one cube of 
planetary masses; legends are shown in upper right of the figure.
In the color version, the green, blue and red lines represent 
planets with their main component is rock, ice and gas, respectively.
}
  \label{fig:ea7}
\end{figure}

In order to explore the dependence of various model parameters, we 
simulated a series of models with a range of different $C_1$ and 
$f_{d,0}$ values.  We use the same values (as Model 1) for [Fe/H]$=0$, 
$\tau_{\rm dep} = 3 \times 10^6$yrs, and $M_* = 1 M_{\odot}$. The 
asymptotic distributions of mass, semimajor axis and eccentricity 
of planets for these models are summarized in Figure~\ref{fig:ea7}.

For high values of $C_1 (= 1)$, type I migration is so efficient 
that gas giants are rarely formed. In runs with $\log f_{d,0} \ga 0$, 
rocky/icy planets are not retained at $a \la 1$AU except the vicinity 
of the disk edge at which type I migration is halted.

\begin{figure}[btp]
\epsscale{1.0}       
\plotone{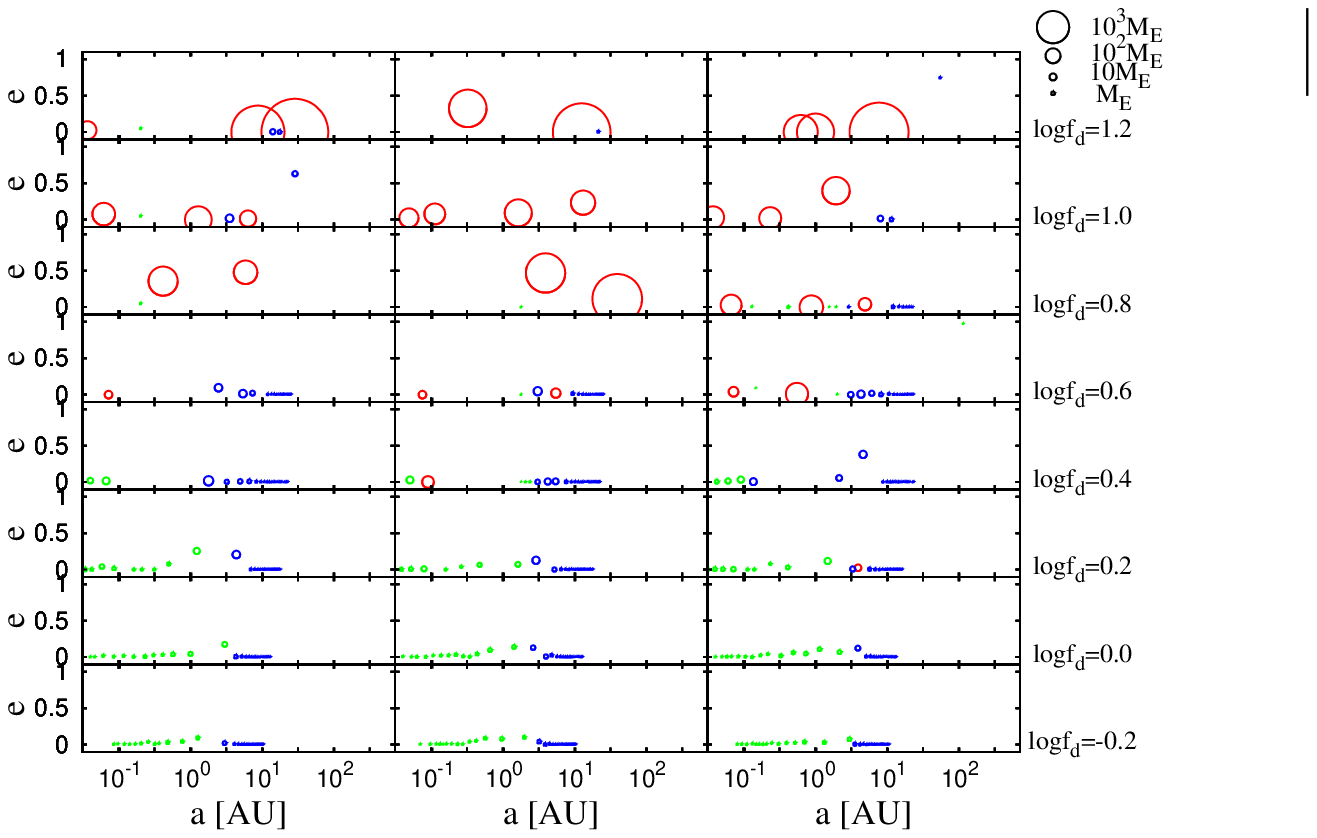} 
\caption{   
Planetary systems with $C_1=0.1$ as functions of $f_{d,0}$.
In order to test the dispersion of the simulated results, 
we choose three random number seeds for the scattering 
prescription.  Models in the middle column are identical 
to those in the middle column in Figure~\ref{fig:ea7}.
}
  \label{fig:ea7b}
\end{figure}

For models with $C_1 = 0.1$ and 0.01, the planet distributions 
are sensitively determined by the total planetesimal mass in the
disk, i.e., the value of $\log f_{d,0}$.  In our procedures for 
close scattering (see \S2.2) depend on the choice of random number 
seeds.  In order to test the robustness and dispersion of our 
models, two additional series of models are carried out for
$C_1 = 0.1$ with different random number seeds for the scattering 
prescription (Figure~\ref{fig:ea7b}).

All these models show similar planetary systems' dependence 
on values of $f_{d,0}$. With relatively high values of $f_{d,0}$, 
the emerging planetary systems consist mostly of gas giants.  In 
these massive disks, the amount of rocky/icy materials is 
sufficiently large to enable the formation of multiple cores 
of gas giants.  Dynamical instability leads to giant planets to
undergo close encounters which excite their eccentricity.  Some
giant planets may collide and coalesce into massive and 
eccentric merger products \citep{Lin_Ida97}. Other planets 
attain eccentric orbits including those which are ejected 
from the system or scattered to endure close encounters with 
their host stars.  During these violent dynamical relaxation, 
the orbits of rocky/icy planets are strongly perturbed by 
the giant planets and they are generally not retained.

In contrast, disks with moderate masses (or values of $f_{d,0}$) 
generally nurture fewer and more widely separated cores.  Gas 
accretion leads to the emergence of either single or widely 
separated and sparsely populated giant planets.  The mass
of these planets is determined by the gap formation condition
and is generally smaller than that of merger products.  
The mutual interaction between single gas giant planets and residual
embryos or between sparsely populated gas giant planets is generally 
too weak for them to become dynamical unstable.  They then
retain relatively low eccentricities.  

Comparison of Figures~\ref{fig:ea7} and \ref{fig:ea7b} 
indicates relatively massive and eccentric gas giant planets are
formed out of massive disks whereas disks with modest masses
generally produce gas giant planets with lower masses and 
eccentricities.  Through the population synthesis models 
(next section), we show that this correlation can be 
used to infer initial formation condition of mature planets 
long after their natal disks have been depleted.  

\section{Population Synthesis of Planetary Systems}

This set of prescription enable us to incorporate 
gravitational interactions among planets into our
population synthesis models.  The main objective
of simulating "multi-planets-in-a-disk" models
is to generate mass and eccentricity distributions 
of the emerging multiple-planet systems.  These 
quantities can be directly compared with 
observational data.

\subsection{Initial conditions}

We adopt a range of disk model parameters (see \S3.1)
and assign each their values with appropriate 
statistical weights to match the observed distribution 
of disk properties.  We assume that $f_{g,0}$ and 
$\tau_{\rm dep}$ have log normal distributions in ranges of
0.1--10 and $10^6$--$10^7$ years, respectively.
We use a normal distribution of [Fe/H] in a range of -0.2 
to 0.2 which corresponds to that of typical target stars 
in various on-going radial velocity surveys. Since the 
target stars of radial velocity surveys are mostly G dwarfs,
we use a log normal distribution, in a range of 
0.8-$1.25M_{\odot}$, for the host stars' mass ($M_*$).

The initial distributions of seed planetesimals are
described in \S3.1 and their growth rate and migration time
scales are specified in \S 3.2--3.5.  Prescriptions for 
dynamical interactions among multiple planets are presented 
in \S 2.1--2.3 and 3.6.

\subsection{Distributions of mass, semimajor axis and eccentricities}

\begin{figure}[btp]
\epsscale{1.0}       
\plotone{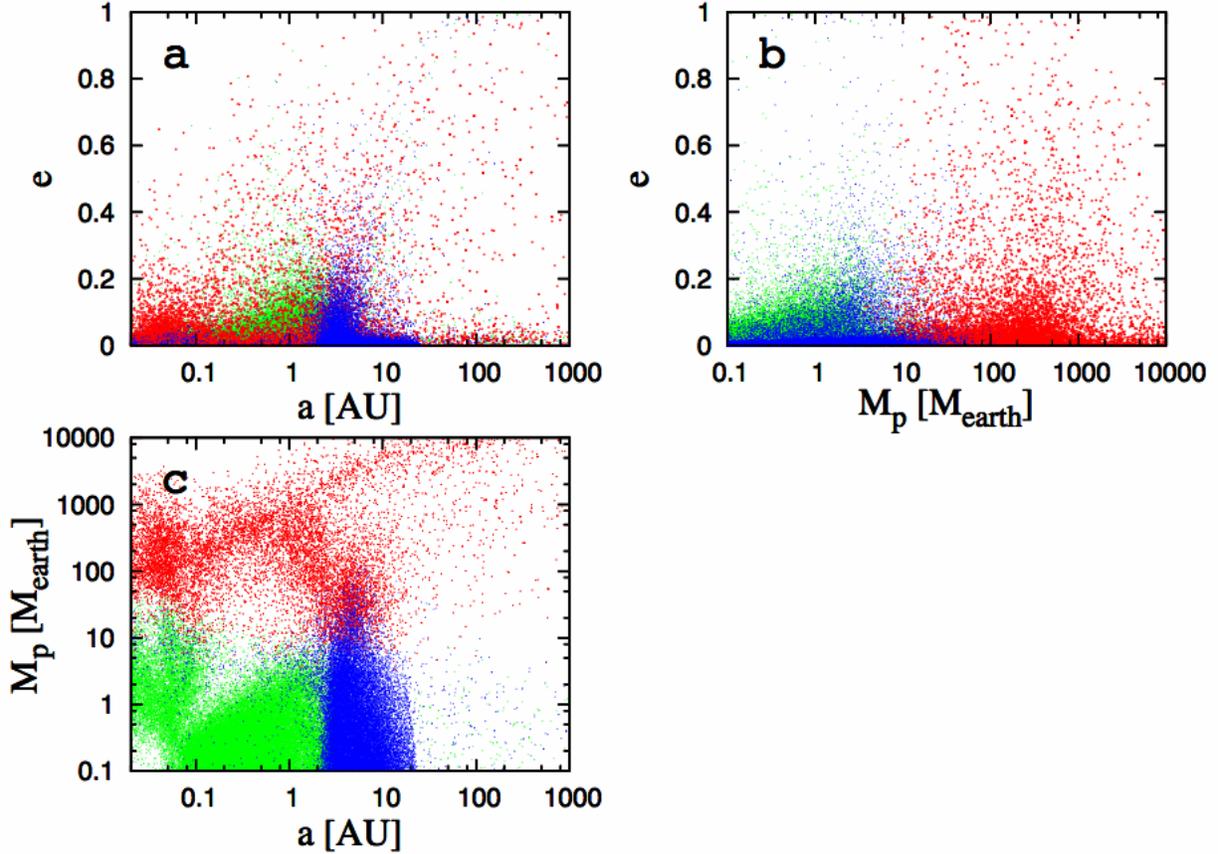} 
\caption{   
The asymptotic kinematic distributions of a simulated population of 
planets in multiple-planet systems.  For these fiducial models, we adopt 
$\tau_{KH 1} = 10^9$ yr and $k_2 = 3$ and $C_1 = 0.1$.  The 
eccentricity $e$ versus semimajor axis $a$ distribution is plotted
in panel (a).  The $e-M_p$ distribution is plotted in panel (b). 
The $M_p-a$ distribution is plotted in panel (c).
The green, blue and red symbols represent 
rocky, icy and gas giant planets, respectively.
}
\label{fig:mea}
\end{figure}

The asymptotic $e-a$, $e-M_p$, and $M_p-a$ distributions of a 
population of $10^4$ multiple-planet systems are shown in 
Figures~\ref{fig:mea}.  Similar to previous figures, the green, 
blue and red symbols represent rocky, icy and gas giant planets, 
respectively.

Models with similar parameters have been simulated in Paper IV
with the "one-planet-in-a-disk" prescription.  In these previous
models, the formation of gas giant planets require a substantial 
reduction in the pace of their cores' type I migration (with 
$C_1 = 0.01-0.03$).  This result reflects the requirement for
not only the formation but also retention of sufficiently 
massive cores (\S3.6) to efficiently accrete gas. The present 
prescription bypasses this competition between type I migration 
and mass growth because it takes into consideration the 
possibility that multiple embryos may undergo convergent migration,
dynamical instability, orbit crossing and close encounters. 
Coalescence of two or more embryos (with relatively low masses and 
long type I migration time scales) promote sporadic, substantial
mass increases.  The merger products can impulsively acquire 
a critical mass for the onset of gas accretion before they 
undergo extensive migration.  

A direct comparison between Figure~\ref{fig:mea}c and Figure 5 
in Paper IV show that, with $C_1=0.1$, the new prescription 
indeed leads to much more prolific production of gas giants.
Some of these gas giant planets form during the advanced stages of
disk evolution after 90\% of the disk gas been depleted.
In these low density environment, the eccentricity of 
embryos may be excited by their perturbation on each other 
but not effectively damped. Eventually they cross each other's
orbit, merge into sufficiently massive cores to accrete
residual disk gas and evolve into gas giant planets.  

In contrast to the ``one-planet-in-a-disk'' models, we can
determine, with the new prescription, planets' velocity dispersion 
from their mutual interaction in multiple-planet systems.  
From this kinematic information, we obtain the $e-a$ and $e-M_p$ 
distributions in Figure~\ref{fig:mea}a and c.  The "V" shape 
centered at a few AU in the $e-a$ distribution is a signature of
strong scatterings by the most massive planets in the system which 
preferentially form near the snow line (at a few AU's).  
Less massive planets are either scattered inward, outward, or 
collide with the gas giant planets.  The left/right branch of 
the ``V'' corresponds to the loci of planets which have been 
scattered inward/outward. These inwardly/outwardly scattered 
planets preserve their apo/peri astrons to be orbital radii
of the gas giant planets.  
Finally, the $e-M_p$ distribution shows that
eccentricity distribution is uniform or rather increases with $M_p$.
If we consider only massive planets that are currently observable
by radial velocity survey, eccentricity distribution clearly 
increases with $M_p$ as shown below.

\begin{figure}[btp]
\epsscale{1.0}       
\plotone{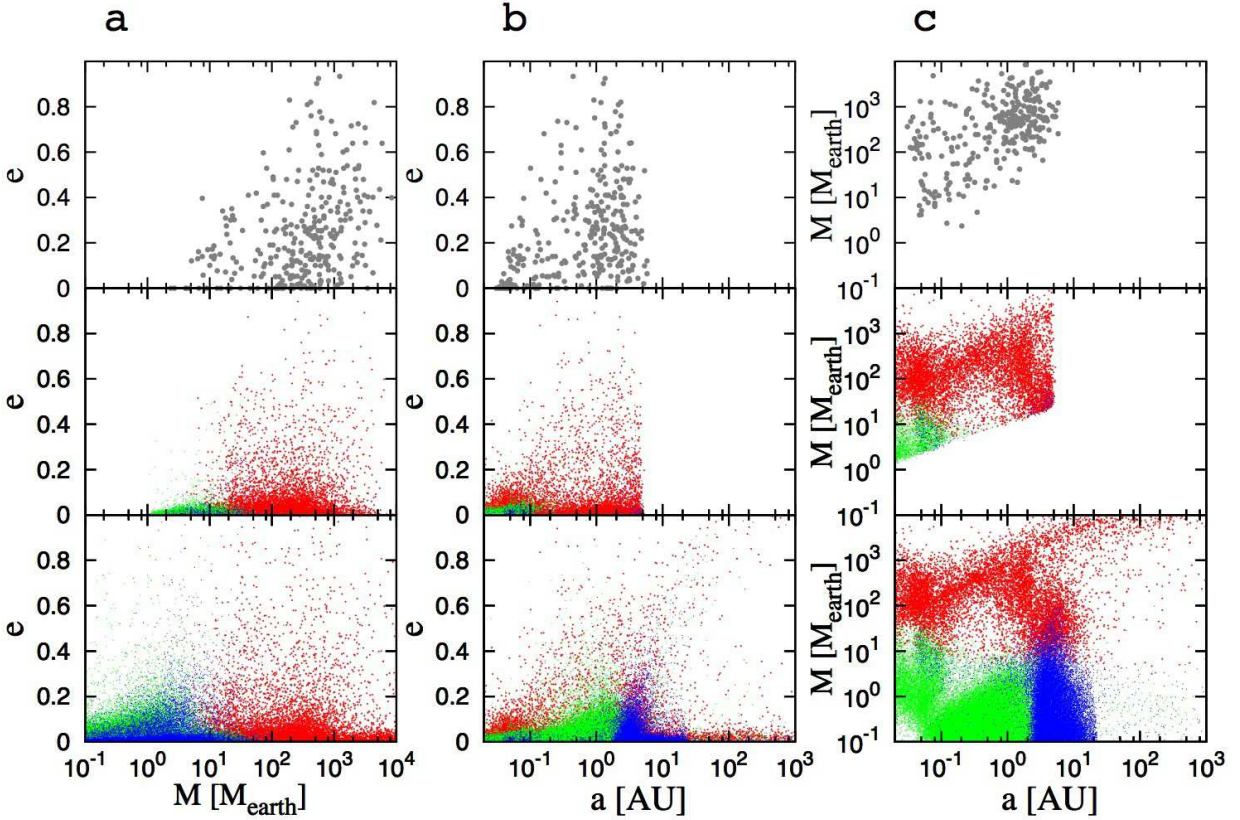} 
\caption{   
Comparison of theoretical results in a nominal case with 
observational data: (a) $e-M_p$, (b) $e-a$, and (c) $M_p-a$
distributions.
The top row are the data obtained from radial velocity surveys.
The planets that were discovered transit survey and confirmed with 
follow-up radial velocity observation are excluded.
The bottom row are the distributions of planets in $10^4$ systems 
simulated with the updated population-synthesis prescription.    
Panels in the middle row represent only the simulated observable 
planets, i.e., those with radial velocities greater than 1 m/s and 
periods less than 10 years.
}
  \label{fig:mea_comp_obs}
\end{figure}

We compare our simulated population-synthesis models with observational
data.  In Figures~\ref{fig:mea_comp_obs}, we plot observed (a) $e-M_p$, 
(b) $e-a$, and (c) $M_p-a$ distributions in the left, middle, and right 
columns respectively.  The observational data are displayed in the top
row.  These data include only those planets which were first discovered
by radial velocity surveys.  We excluded planetary candidates which were
first discovered by transit surveys, regardless whether they were confirmed
by follow-up radial velocity observations.  The main rationale for this
selection is that the detection probability by transit surveys is highly 
biased to short-period planets.

The simulated population of $10^4$ multiple-planet systems is plotted
in the bottom row.  Here $M$ refers to $M_p \sin i$ with a 
distribution of inclination angle $i$ between the line of sight and the planets'
orbital angular momentum axis that assumes random orientations of
orbit normals of the planets.
Some of these planets have low-mass and large
semimajor axis.  They are below the present-day detection capability.  
In order to take into account the limited radial velocity precision 
and data base line, we selected a subset of data in which the predicted
reflective motion of the host stars is larger than 1 m/s and the 
planets' orbital period is shorter than 10 years.  These ``detectable
planets'' are plotted in the middle row.  

Comparison between top and middle panels in 
Figures~\ref{fig:mea_comp_obs}a and b shows that the theoretical 
models match well the observed kinematic distributions.
Both observations and population synthesis models show
that $e$ increases with $M_p$ and $a$.
In the next subsection, we will show that these correlations
are robustly reproduced, almost independent of model parameters.

Most gas giants have semimajor axis $a \ga 0.5$AU.  Their 
eccentricities are not affected by the dissipation of tidal
perturbation induced by the host stars on the planets.  
Detailed comparison of the top and mid row of the middle
column indicate that the fraction of high-eccentricity 
($e \ga 0.2$) planets in the theoretical model is low
($\sim 10\%$) compared with that ($\sim 50\%$) in observed data.
This discrepancy may be observational uncertainties which generally
tend introduce over-estimates for planets' eccentricities.  
It may also be due the secular perturbations between gas giants 
which has not yet been incorporated in our prescription.  In 
multiple-planet systems with significant angular momentum deficit,
secular interaction can induce large-amplitude angular momentum 
exchange, eccentricity modulation, and long-term orbital instabilities.
A treatment of secular perturbations will be 
constructed and presented elsewhere.
 
In this paper, we focus primarily on the planets' eccentricity distributions.
Comparison between top and middle panels in Figure~\ref{fig:mea_comp_obs}c
show that our prescription qualitatively reproduces the observed correlations
in the $M_p-a$ distribution.  Whereas the observational data clearly show 
an over density of gas giants at $a \ga 0.7$AU, their semimajor axis 
distribution in the simulated population synthesis models is relatively 
uniform on a log scale.  The sensitive dependence of the $M_p-a$ 
distribution on the parameters of population synthesis models will be 
discussed in detail in the next paper.

\subsection{Correlations between eccentricity and mass or semimajor axis}

The origin of the correlation of eccentricity and semimajor axis can be
traced back through the procedures of our theoretical model. In the observed 
$e-a$ distribution, $e$ is apparently lower at $a \la 0.1$AU than at
 $a \sim 1$AU (the top panel in Figure~\ref{fig:mea_comp_obs}b).  This correlation
has been attributed to the dissipation of stellar tidal perturbation
inside the planetary envelope which leads to the orbital circularization 
of close-in planets.  This tidal effects has not yet been implemented in
the population synthesis models.  Yet, this $e-a$ correlation is also
very well established in the simulated results.  In Eq.~(\ref{eq:e_escG0}),
we showed that the maximum eccentricity excited by close scattering between
gas giants is $\propto a^{1/2}$, because the two-body surface escape 
velocity ($v_{{\rm esc},jk}$) is independent of $a$ whereas the Kepler 
velocity ($v_{\rm K}$) is $\propto a^{-1/2}$.  Consequently, in the deep 
potential near the host stars, it is difficult for close scattering to 
excite high eccentricities. Based on the above prescription, the maximum 
$e$ we obtain at $a \sim 0.1$AU is 3 times smaller than that at 
$a \sim 1$AU in the simulated models  (the middle panel in 
Figure~\ref{fig:mea_comp_obs}b).

\begin{figure}[btp]
\epsscale{1.0}       
\plotone{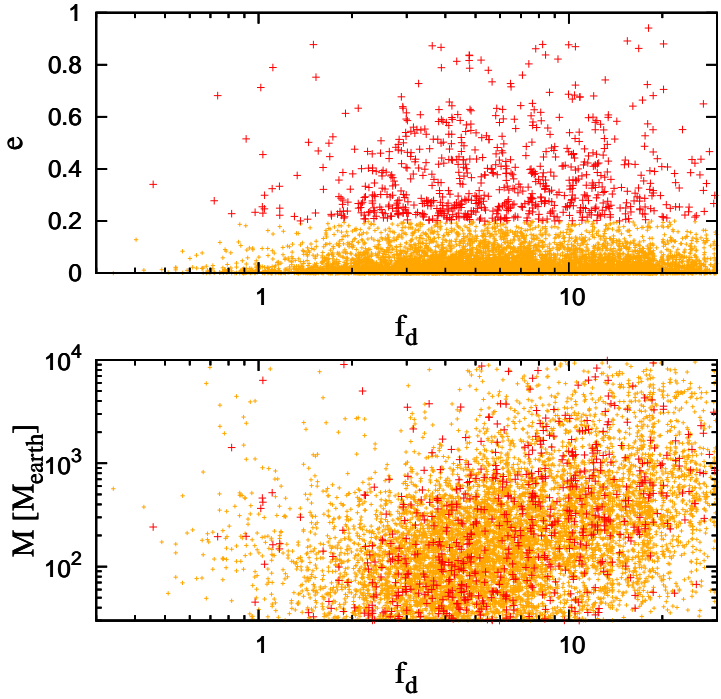} 
\caption{   
The distributions of $e$ and $M$ of gas giants with $a < 30$AU
as a function of $f_{d,0}$ in the fiducial models of Figure~\ref{fig:mea}.
The thick and thin symbols represent giants with $e>0.2$ and
$e<0.2$, respectively (In the color version, red and orange symbols).
}
  \label{fig:fd}
\end{figure}

Multiple massive giants are preferentially formed in relatively 
massive disks (see discussions in \S4). These dynamically packed
systems are more prone to dynamical instabilities, orbit crossing,
excitation of high eccentricities and cohesive collisions.  
A correlation between high eccentricity and planetary mass is 
naturally expected. In order to verify this scenario, we plot in 
Figure~\ref{fig:fd}, $e$ and $M_p$ of gas giants as a function 
of $f_{d,0}$ in the fiducial models (in which $\tau_{KH 1} = 
10^9$ yr and $C_1 =0.1$) of Figure~\ref{fig:mea}. This
figure clearly shows that, in the range of $f_{d,0} = 1-10$, both 
the mass of gas giants and the fraction of high $e$ planets 
increase with $f_{d,0}$.  Based on this inference, we can infer
the distribution of disk masses from the observed $e-M_p$ distribution.
\citet{Raymond10} performed N-body simulations of multiiple
giant planets with different masses and found a similar correlation 
between high eccentricity and planetary mass,
although their initial conditions are artificial.
Note that Figure~\ref{fig:fd} only includes gas giants at $a < 30$AU.
As shown in Figures~\ref{fig:mea}a and c, the kinematic distribution 
of gas giants with $a \ga 30$AU have two (high $e$ and low $e$) components.
We discuss this issue in the next subsection.

\begin{figure}[btp]
\epsscale{1.0}       
\plotone{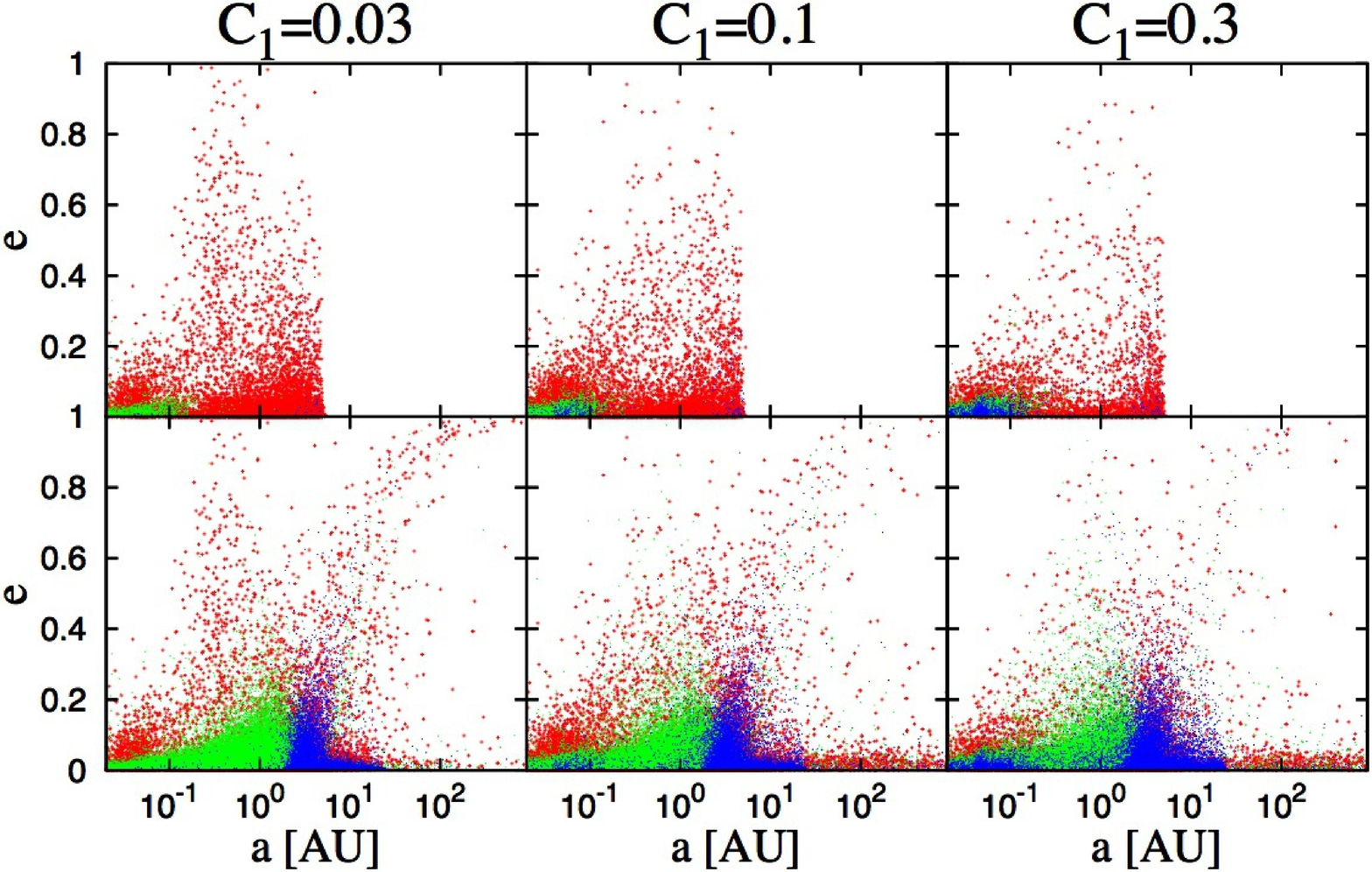} 
\caption{   
Dependence on $C_1$ of the simulated $e-a$ distribution.
The lower panels are distributions of $10^4$ systems of planets
generated with our Monte Carlo simulation.  The upper panels 
contain only the observable (with radial velocities greater than 1 m/s
and periods less than 10 years) simulated planets.  These 
observable criteria correspond to the technical limitation of 
current radial velocity surveys. The left, middle and right panels 
display results for models with type I migration coefficient 
$C_1=0.03, 0.1$ and 0.3, respectively.  The symbols are the 
same as those in Figures~\ref{fig:mea_comp_obs}.
}
  \label{fig:C1ea}
\end{figure}

\begin{figure}[btp]
\epsscale{1.0}       
\plotone{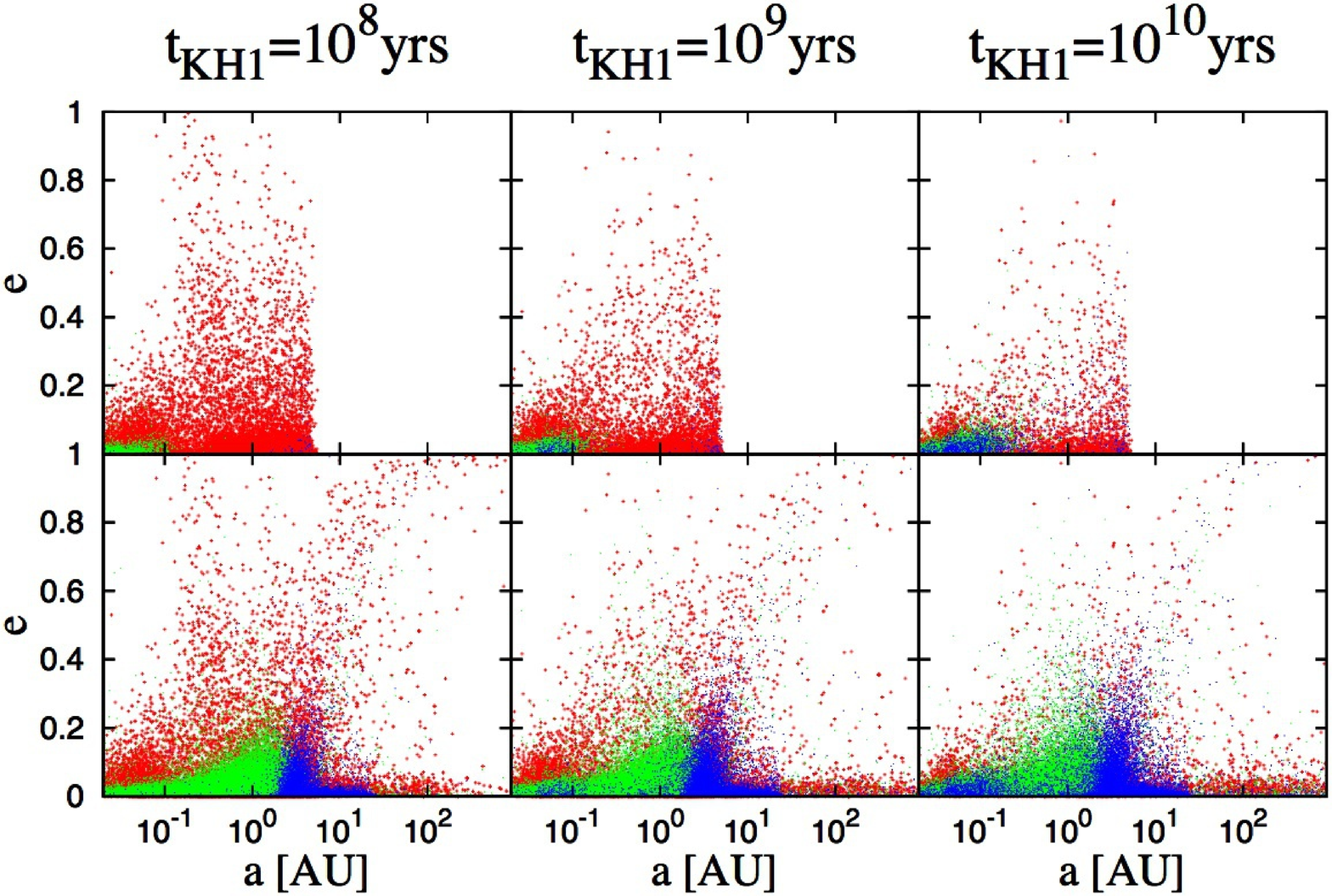} 
\caption{   The $e-a$ distribution of simulated models with 
various values of $\tau_{\rm KH1}$.  The left, middle and 
right panels include results generated with the parameter
for envelope contraction timescale (for $M_p = M_{\oplus}$) 
$\tau_{\rm KH1}=10^8, 10^9$ and $10^{10}$ years, respectively.
The meaning of upper and lower panels are the same as in 
Figures~\ref{fig:C1ea}.
}
  \label{fig:KHea}
\end{figure}

The above discussion indicates that the $e-a$ and $e-M_p$ 
correlations are the natural outcome of close scatterings 
and the formation of gas giants.  It also shows that these 
correlations are insensitive to other physical parameters 
such as the rate of type I migration rate ($C_1$) and the 
magnitude of the gas accretion timescale ($\tau_{\rm KH1}$).  
These independences are reflected in Figures~\ref{fig:C1ea} 
and \ref{fig:KHea} for simulated populations with different
values of $C_1$ and $\tau_{\rm KH1}$.  Although the number of
gas giants per host star is lower among populations with 
relatively high $C_1$ and long $\tau_{\rm KH1}$, the trend
that $e$ is lower at $a \la 0.1$AU than at $a \sim 1$AU
persists.

This trend is less conspicuous in the limit of relatively small 
$C_1$ ($0.03$ in Figures~\ref{fig:C1ea}) and gas accretion time
scale $\tau_{\rm KH1}$ ($10^8$ years in Figure~\ref{fig:KHea}).
In these limits, the prolific formation of gas giants and the
high frequency of their close scattering leads to a prominent
"V" feature in the $e-a$ distribution.  This feature also affects
the distribution of observable planets (displayed in the upper 
rows of these figures).  The lack of any obvious evidences of this
``V'' feature in the observational data (Figure~\ref{fig:mea_comp_obs}a),
places a limit on the efficiency of gas giant formation and 
constraints on the magnitude of $C_1$ and $\tau_{\rm KH1}$.
In the fiducial population synthesis model (with $C_1 =0.1$
and $\tau_{\rm KH1}=10^9$ yr), the fraction of solar-type stars 
that harbor gas giants with $a=0.5-5$AU and $M_p > 100M_{\oplus}$ 
(cool jupiters) is 21\%.  This fraction for both the $C_1=0.03$
models (47\%) and the $\tau_{\rm KH1}=10^8$ years models (42\%)
are both much higher than that inferred by radial velocity surveys.

\begin{figure}[btp]
\epsscale{1.0}       
\plotone{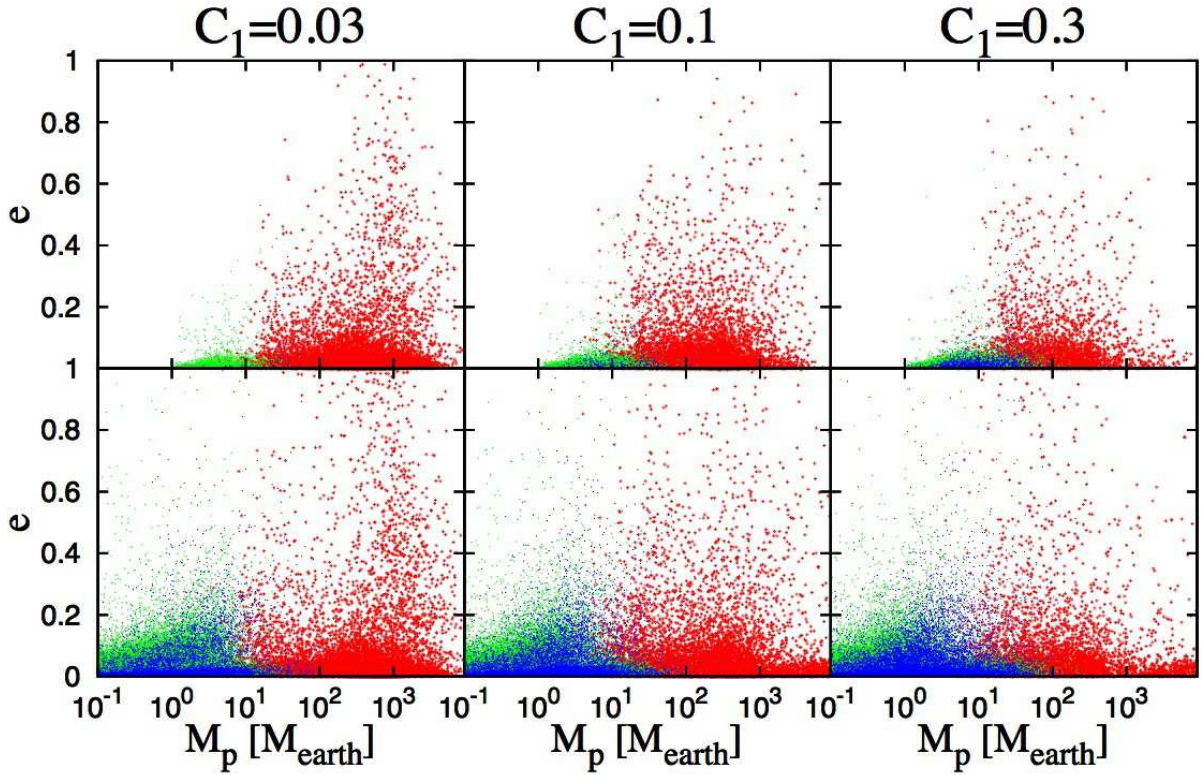} 
\caption{   
Dependence on $C_1$ of the predicted $e-M_p$ distribution.
The meaning of panels are the same as in Figures~\ref{fig:C1ea}.
}
  \label{fig:C1em}
\end{figure}

\begin{figure}[btp]
\epsscale{1.0}       
\plotone{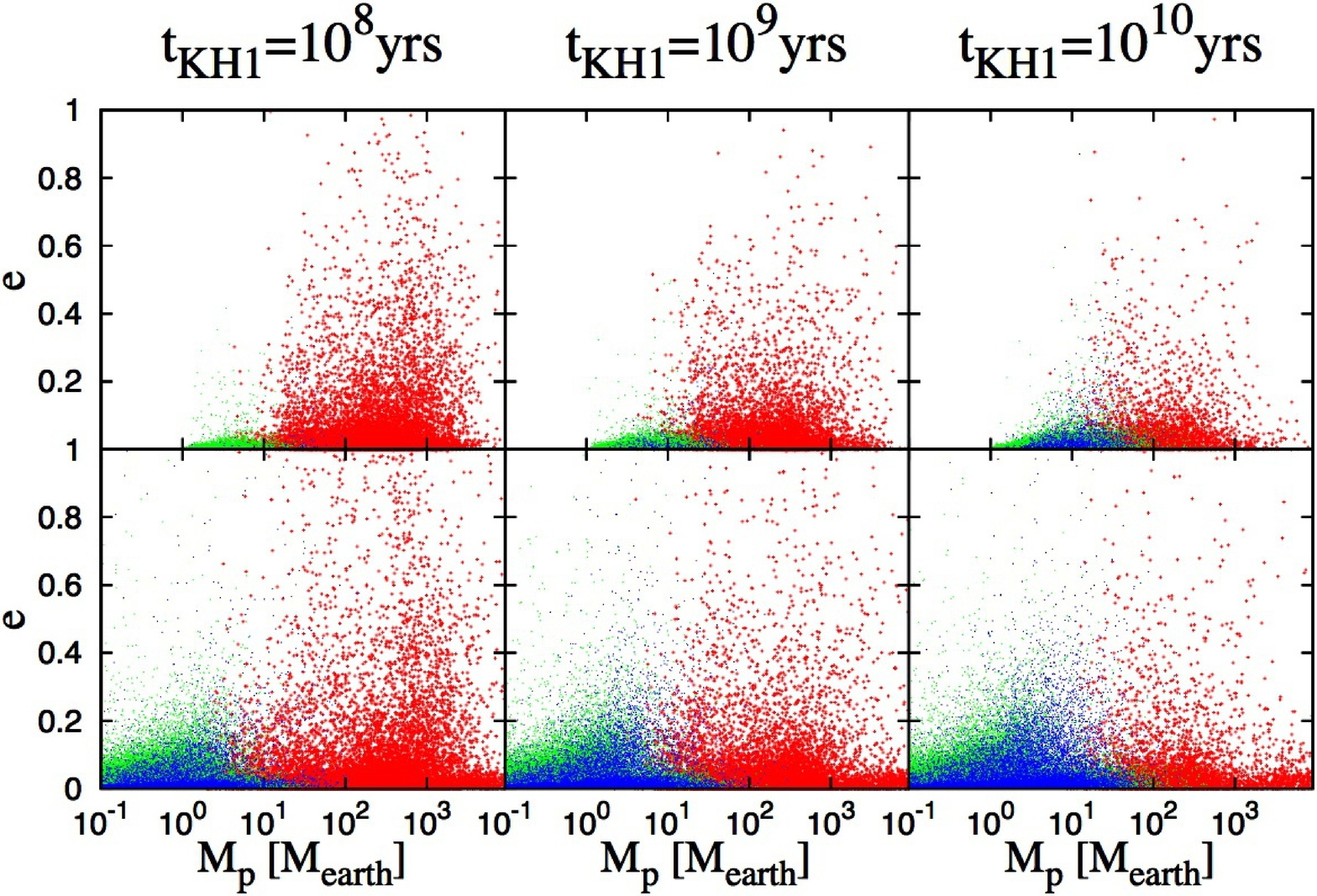} 
\caption{   
Dependence on $\tau_{\rm KH1}$ of the predicted $e-M_p$ distribution.
The meaning of panels are the same as in Figures~\ref{fig:KHea}.
}
  \label{fig:KHem}
\end{figure}

Figures~\ref{fig:C1em} and \ref{fig:KHem} show 
the dependence of the $e-M_p$ distribution
on $C_1$ and $\tau_{\rm KH1}$, respectively.
Although more massive planets, in general, tend to 
have larger $e$, this trend is less conspicuous 
for models with $C_1=0.3$ and $\tau_{\rm KH1}=10^{10}$ years.
In these limiting cases, the formation of gas giants 
is affected by the rapid migration of the cores 
and slow gas accretion.  These restrictions are more severe 
for formation of the relatively massive gas giants.
Despite these extreme cases, the correlations in $e-a$ 
and $e-M_p$ distributions are well established in the 
population synthesis models for a wide range of parameters.  
The general reproduction of the observed correlations 
suggests that these properties are associated with
close scattering among gas giants.

\begin{figure}[btp]
\epsscale{1.0}       
\plotone{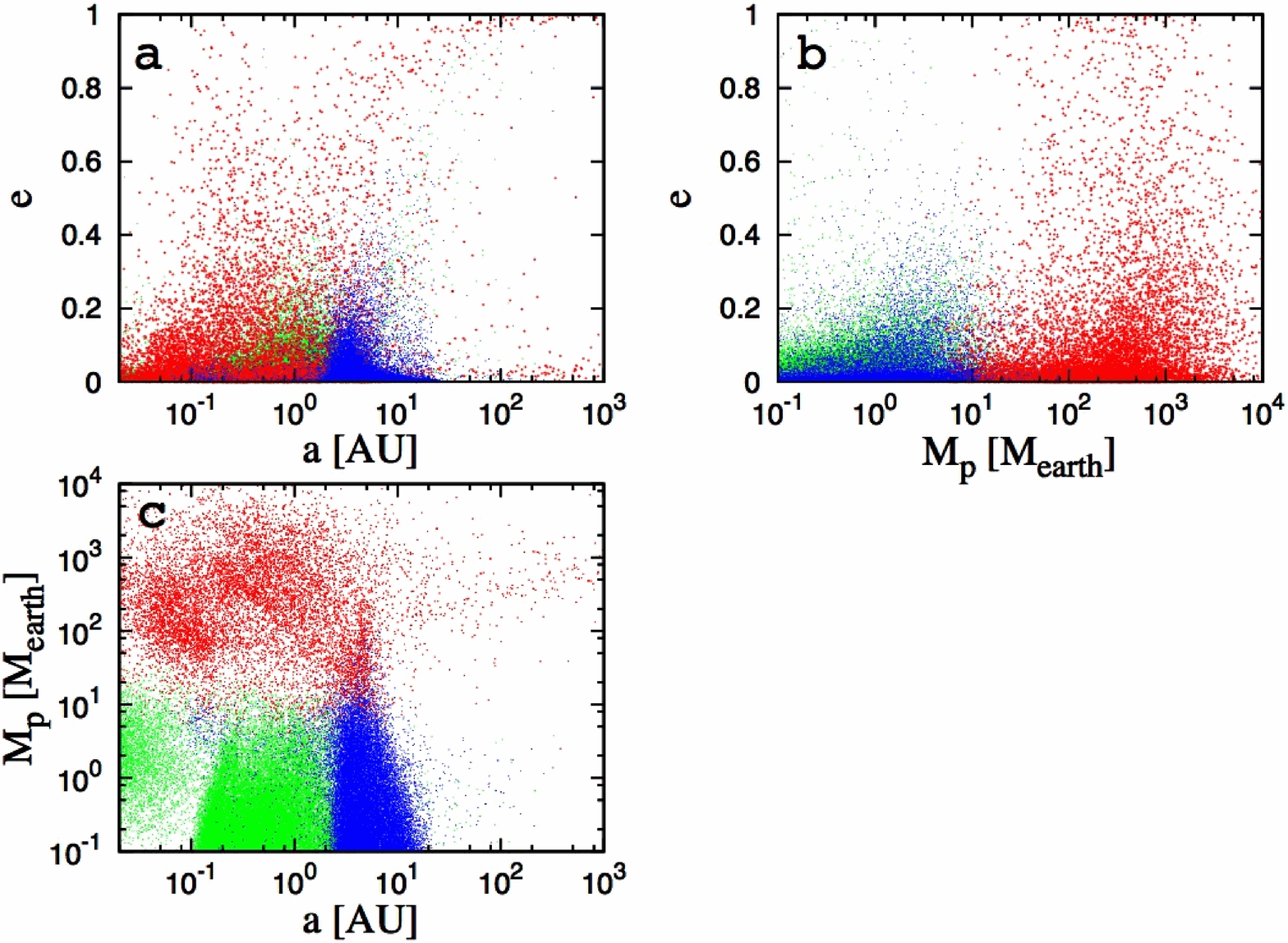} 
\caption{   
Same as Figure~\ref{fig:mea} except for 
use of non-isothermal type I migration formula 
instead of isothermal one with $C_1 = 0.1$.  }
\label{fig:meaP}
\end{figure}

In Figures~\ref{fig:meaP}a, b and c, we 
show the $e-a$, $e-M_p$ and
$M_p-a$ distributions with the non-isothermal
type I migration formula. 
Because in inner (optically thick) disk regions,
migration is outward  for some range of $M_p$ \citep{Kretke12}, 
many gas giants
survive even without any reduction in amplitude of the migration velocity.
While the $M_p-a$ distribution is affected by the different
migration formula, the $e-a$ and $e-M_p$ distributions
are similar to those obtained by the isothermal migration formula
and they are within a range of variations of the results with 
different values of $C_1$ and $\tau_{\rm KH1}$ 
in figures ~\ref{fig:C1ea}, \ref{fig:KHea}, \ref{fig:C1em}
and \ref{fig:KHem}.
Therefore, here we do not go into the details of 
the non-isothermal migration formula and
focus on the eccentricity distributions obtained with
the isothermal migration formula and a scaling factor $C_1$,
as discussed in \S 3.3.

\subsection{Formation of distant large gas giants in circular orbits}

The population synthesis models also generate a population of massive
($M_p \ga 10M_J$) gas giants with large semimajor axis  ($a \ga 30$AU) 
(Figure~\ref{fig:mea}c). In principle, these planets can be observationally
detected during their infancy with direct imaging searches.  In the 
fiducial models, the fraction of host stars that have these planets 
is $\sim 3.7\%$ and most of them have low-eccentricity ($e \la 0.1$) 
orbits (Figure~\ref{fig:mea}a).  If these planet were grown to gas giants 
near the 
snow line and scattered to these large distances, they would preserve
their periastron distance.  For example, planets formed interior to 
30 AU may be scattered to attain a semimajor axis $ga 100$ AU with 
an $e \ga 0.7$.  Although the effect of eccentricity damping is included
in our prescription for disk-planet interaction, its efficiency is generally
too low to account for the low eccentricity of distant, massive planets
generated in these population synthesis models (step 3-a of Appendix A1).

A close inspection of our results yields an alternative path for the
formation of distant gas giant planets.  In systems which contain 
multiple gas giants, the rapid gas accretion of first generation gas
giants destabilize the orbits of nearby residual embryos.  Some embryos
are scattered to large distance where the surface densities of both residual
planetesimals and gas are relatively low.  
Because the scattered embryo masses are usually well below the
local disk mass, even the relatively low surface density gas damps
the eccentricity of the scattered embryos to the level below 0.1
(also see the discussion below).

A reduction in the planetesimal
accretion rate at these distance also lowers the critical core mass for 
the onset of gas accretion (see  Eq.~\ref{eq:crit_core_mass}). In
extended protostellar disks, there is adequate supply of gas supply,
even during the advanced stages of disk evolution to enable
the circularized planets to acquire masses larger than that of Jupiter.  

The ratio of 
planets' asymptotic mass to that of their host stars is a monotonically
increasing function of their natal disks' aspect ratio which is also
an increasing function of $a$ (see Eqs.~\ref{eq:m_gas_vis} and 
\ref{eq:m_gas_th}).  The simulated populations confirm that the 
masses of the long-period gas giants are robustly correlated
with $a$ in the fiducial (Figure~\ref{fig:mea}c) and other models.
If this trend is confirmed by observation, it provides evidence to
support the scenario that long-period, low-eccentricity, massive gas 
giants are the byproducts of gas accretion onto outwardly 
scattered cores.  

During the course of mass accretion, the planets also acquire 
angular momentum of the disk gas.  Although the scattered embryos
have high eccentricity initially, they gain angular momentum through 
planet-disk tidal interaction.  Our prescription for the eccentricity
damping may be less efficient than that, due to supersonic relative 
velocity between the core and disk gas \citep{Ostriker99, 
Papaloizou00, Muto10}, although it nonetheless may lead to the circularization of
planets' orbits.  In addition to the tidal torque, E. Thommes (2010, 
private communication) used a hybrid N-body and 2D hydrodynamical (FARGO) 
scheme to demonstrate that during their eccentric excursion, 
scattered embryos
take longer time to pass through their apoastrons and they tend to 
accrete gas from that region which has relatively large specific 
angular momentum. 
We anticipate that as the embryos evolve 
into gas giants, their specific angular momentum becomes comparable 
to that of the disk gas near their apoastron and their orbits become 
circularized with a radius close to their apoastron radius.
The detailed calculation will be shown in a separate paper
(Kikuchi, Higuchi \& Ida, in preparation). 

Through this process, multiple cores can be scattered outward 
to initiate gas accretion.  We find that in the fiducial models,
the fraction of systems with two, three, and four giants with
low-eccentricity orbits (such as the solar system) is 0.4\%, 
0.07\%, and 0.04\%, respectively. The orbital configuration
of systems of multiple gas giants with relatively large masses, 
low eccentricities, and long periods are consistent with the
observed data obtained from the direct imaging initiatives.
 
We have not introduced an outer boundary for planets' natal disks.
If most protostellar disks are truncated either by photoevaporation
or magnetic braking, the outwardly scattered cores would not have 
access to gas beyond the truncation radius. An abrupt decline in
the observed same major axis distribution of gas giant planets 
could be used to place constraints on the structure and evolution 
of their natal disks. 

In the models with $C_1=0.3$, the fraction of systems with 
massive, low-eccentricity, and long-period gas giants 
is similar to that of the fiducial models (with $C_1=0.1$).
But this fraction is reduced to $0.04\%$ in the models 
with $C_1=0.03$.  This dichotomy suggests that type I 
migration of cores is a major cause for them to be 
scattered by a giant planet. In order to match with 
the frequency of planets discovered through direct 
imaging observation, type I migration cannot be neglected.

\section{Summary and Discussions}
\label{sec:discussions}

Radial velocity and transit surveys indicate that multiple-planet
systems are common around solar-type stars. We adopt a new paradigm
that all planets are formed in families (as in the case of the solar 
system) and their asymptotic properties are determined by their 
dynamical interaction with each other as well as with their natal disks.

In order to incorporate this scenario into our population synthesis 
models, we upgrade an existing numerical population synthesis scheme 
which was developed in Paper I-IV for idealized ``one-planet-in-a-disk''
models.  The original prescriptions were constructed and applied to 
simulate the semimajor axis and mass distributions 
of extrasolar planets, 
calculating the growth and migration of each planet, independently 
of any other planets.  
Similar approaches have also been adopted 
by other authors \citep{Mordasini09,Mordasini09b,Alibert11}. 

In Paper VI, we first introduced a modification to partly include 
the gravitational interactions and collisions between rocky 
planetary embryos.  In this paper, we generalize an efficient and 
robust prescription to take into account close scattering between 
all planets.  We calibrate our prescriptions for scattering among 
gas giants with the results of comprehensive N-body simulations. 
For example, in the aftermass of close scatterings between gas 
giants, ejection commonly occurs.  Our prescriptions reproduce 
well the distributions of eccentricity and semimajor distributions 
of retained giants obtained from the N-body simulations.

With this upgraded tool, we can adopt a Monte-Carlo approach to
simulate a large sample of population synthesis models.  In this
paper, we mostly investigate the effects of scattering by giant 
planets on asymptotic architecture of planetary systems and 
their eccentricity distributions.  We focus on the correlations 
between the eccentricity of gas giants with their mass and 
semimajor axis.

Observational data obtained through radial velocity surveys 
clearly shows that the mean eccentricity increases with planetary 
mass (at least for $M \la 10M_J$) and with semimajor axis
(at least for $a \la 1$AU). Our new population synthesis models 
show that in relatively massive disks, several massive giants 
may form and interact with each other intensely.  Dynamical
instabilities lead to eccentricity excitation, orbit crossing,
close encounters, collisions, and ejection. But in protostellar
disks with moderate masses (comparable to the surface density
distribution of the minimum mass nebula model), one or two 
relatively small-mass giants are formed. They commonly avoid 
dynamical instability and preserve their low initial eccentricities.
The observationally inferred correlation between gas giants' eccentricity 
and mass is robustly reproduced by our models.  

Both gas giants and rocky/icy planets form prolifically in massive
disks.  But the low-mass planets are mostly ejected or accreted by the
gas giants. 
Moderate planetary systems in which rocky/icy planets
preserve their modest eccentricity and coexist with one or two 
sibling gas giants are preferentially formed in disks with modest masses
(analogous to the formation of the solar system in a minimum mass solar
nebula).  
Multiple, short-period super-Earths, which are commonly found 
around solar-type stars, are also retained only in the systems formed from 
such modest-mass disks but not in the massive disks. 

The observed correlation between eccentricity and semimajor axis is
also reproduced by the population synthesis models.  After the onset
of dynamical instability and orbit crossing, planets' velocity 
dispersion is determined by close scatterings.  The maximum 
recoil speed is comparable to the surface escape velocity of 
the dominant perturber, which is independent of semimajor axis.
Since mean eccentricity is given approximately by velocity 
dispersion scaled by the local Keplerian velocity, the mean 
eccentricity is proportional to square root of semimajor axis.
Although we have not included the effect of tidal circularization, 
which preferentially damps the eccentricity of short-period 
planets, the correlation of eccentricity and semimajor axis can 
also be produced by close scatterings alone.

Recently, direct imaging surveys discovered several
systems which contain massive ($M_p \ga 10M_J$) gas giants
in distant ($a \ga 100$AU) orbits with small eccentricities ($e\la 0.1$).
Some of these planets are members of multiple planets systems.
We found analogous planets among a non-negligible fraction (a few \%) 
of systems in our population synthesis simulations. In our models,
the cores of these gas giants were originally formed well interior 
to their present-day semimajor axis (with $a \la 30$AU).  They were 
scattered outwardly by first generation gas giants to the outer 
regions of their natal disks.  The surface density of field 
planetesimals and their collision frequency with the cores 
rapidly decreases with the disk radius.  A reduction in the 
planetesimal bombardment rate also decreases the critical core mass 
needed for the onset of efficient gas accretion. 
In the outer regions of relative massive disks, there is an 
adequate supply of gas to enable the cores to accrete gas even 
during the advanced stage of disk evolution. The high eccentricity 
of the scattered cores is damped by disk-planet interactions and 
accretion of gas in the outer disk where its specific angular 
momentum is relatively high. 

The emergence of single massive giants can scatter 
multiple nearby cores, we also found systems with 
multiple massive gas giants in nearly circular distant orbits.
Therefore, the core accretion model can indeed produce the 
systems that are similar to those discovered by direct imaging 
surveys.  This origin predicts a clear correlation that the 
mass increases with semimajor axis for these giants
by the following reason.
The aspect ratio of the 
disks increases with their radius.  At these large distances, gas 
giants continue to accrete gas until they acquire sufficient mass
to open gaps in the disks near their orbits. Because the critical 
mass for gap opening is higher for larger orbital radius, 
the correlation is established.
It is of interest to check if observed data also indeed
exhibits such a trend.

\vspace{1em} 
\noindent ACKNOWLEDGMENTS.  This work is partially supported by NASA
(NNX08AM84G), NSF (AST-0908807), JSPS (23103005) and 
a University of California Lab Fee grant.

\vspace{1em}
\noindent CORRESPONDENCE should be addressed to S. I. (ida@geo.titech.ac.jp).

\clearpage

\section*{Appendix. Prescription for eccentricity excitation and
ejection of giant planets as a result of orbital instability}

In our simulations, giant planets and rocky/icy planetary embryos 
co-exist. We separate the treatment of planets' dynamical 
interactions into those between (i) giant planets, (ii) embryos, 
and (iii) a giant planet and an embryo.

We consider two-giants scatterings in a disk environment with residual 
gas. We adopt an assumption that three-giants scatterings mostly occur 
after the disk has been depleted in a gas-free environment (see \S2).
The prescriptions for scattering between two gas giants and three 
gas giants are described in A1 and A2 respectively.  In A3, we 
present prescriptions for dynamical interactions in general cases 
in which giant planets and rocky/icy planetary embryos co-exist.

\subsection*{A1. Two giants}

The prescriptions for scattering of two giants with mass
$m_1$ and $m_2$ are as follows:
\begin{description}
\item[1)] {\it Specify initial conditions:} 
We assume that orbital instability between two gas giants occurs 
when their orbital separation becomes smaller than $2\sqrt{3} 
r_{\rm H}$, where $r_{\rm H} = ((m_1+m_2)/3 M_{\ast})^{1/3} a$,
$a=\sqrt{a_1 a_2}$, $m_j$ and $a_j$ are the mass and semimajor 
axis of planets $j$ with $(j=1,2)$.  For comparison with 
N-body simulations, we consider nearly circular and coplanar 
orbits (see \S2.1).

\item[2)] {\it Compute trial eccentricities:}
After close encounters, the maximum eccentricity of body $j$ 
($j=1,2$) is given by
\begin{equation}
e_j^{\rm max} = \frac{m_k {\cal R}_j}{m_1 + m_2} e_{\rm esc,12} ,
\label{eq:e_escj0}
\end{equation}
where 
\begin{equation}
e_{{\rm esc},jk} = \frac{v_{{\rm esc},jk}}{v_{\rm K}}
= \frac{\sqrt{2G(m_j+m_k)/(R_j + R_k)}}{\sqrt{G M_{\ast}/a}}
\simeq 1.6 \left(\frac{m_j+m_k}{M_{\rm J}}\right)^{1/3}
  \left(\frac{\rho}{1{\rm gcm}^{-3}}\right)^{1/6}
  \left(\frac{a}{1{\rm AU}}\right)^{1/2}, 
\label{eq:e_escG}
\end{equation}
$v_{\rm K}$ is the Keplerian velocity, $v_{{\rm esc},jk}$
is the two-body surface escape velocity, 
and $R_j$ is the physical radius of planet $j$. 
In Eq (\ref{eq:e_escj0}), ${\cal R}_j$ is a random number 
chosen from a Rayleigh distribution with the root mean 
square of unity \citep{Ida_Makino92}.
We use different seed random numbers for different bodies.
This is the same expression as we constructed, in Paper VI, 
to simulate eccentricity excitations of embryos.

\item[3-a)] {\it Execute a planet's ejection}
\begin{enumerate}
\item {\it Select the body to be ejected:}
We specify that an ejection would occur if $e_j^{\rm max} > 1$ for 
at least one of the bodies. (Since the total energy of the system
is conserved to its negative values prior to the close encounters, 
it would not be possible for both planets to escape from their host
stars).  We identify the ejected planet with a candidate that has a 
larger value of $e_j^{\rm max}$.

\item {\it Evaluate the asymptotic semimajor axis and eccentricity 
of the retained planet:}
N body simulations indicate that in two planet systems, ejection
of a planet primarily occurs after a series of weak distant 
encounters between it and the retained planet.  As it attains
a parabolic orbit, the eccentricity of the escaping planet 
(with a subscript label $j=2$) gradually increases slightly above unity.  

We assume that the initial (denoted by ",0") orbits of the two 
giants were circular. Angular momentum conservation during the
close encounters would imply
\begin{eqnarray}
m_1 \sqrt{a_{1,0}} + m_2 \sqrt{a_{2,0}} & = &
m_1 \sqrt{a_1 (1+e_1)(1-e_1)} + m_2 \sqrt{a_2 (1-e_2)(1+e_2)} \\
 & \simeq & m_1 \sqrt{a_{1,0} (1-e_1)} + m_2 \sqrt{a_{2,0}\times 2},
\label{eq:ang_momentum}
\end{eqnarray}
where $a_1$ and $e_1$ are the asymptotic semimajor axis 
and eccentricity of the retained gas giant (with a subscript 
label $j=1$).  In the above approximation, we have assumed 
the post-encounter apoastron distance of the retained 
planet $a_1 (1 + e_1) \sim a_{1, 0}$ and periastron distance
of the ejected planet $a_2 (1 - e_2) \sim a_{2, 0}$.

From this equation, we obtain
\begin{eqnarray}
e_1 \simeq e_1^* & \equiv & 2(\sqrt{2}-1) 
\frac{m_2}{m_1}\sqrt{\frac{a_{2,0}}{a_{1,0}}}
- (\sqrt{2}-1)^2 \left(\frac{m_2}{m_1}\right)^2 
\frac{a_{2,0}}{a_{1,0}} \\
  & \simeq & 0.83 \frac{m_2}{m_1} - 0.17 
\left(\frac{m_2}{m_1}\right)^2.
\label{eq:ang_momentum2}
\end{eqnarray}
In order to reproduce the results of N-body simulation,
we assume that close encounters generates $e_1$ with 
a Gaussian distribution with a peak value around 
$\sim e_1^*$ and a dispersion $\sim (e_1^*/3)$:
\begin{equation}
f(e_1)de_1 = \frac{1}{\sqrt{2\pi} (e_1^*/3)} \exp 
\left(-\frac{(e_1 - e_1^*)^2}{2(e_1^*/3)^2} \right) de_1.
\label{eq:e_escjr2}
\end{equation} 
Energy conservation implies that at least one of the two planets must
remain gravitationally bound to the host star after the intense scattering.  
This requirement also implies that $e_1$ must be $< 1$.  We truncate the 
the high value tail in the $e_1$ distribution and renormalize
the numerical factor such that integral of the probability is unity.

After the scattering, disk-planet interactions tend to damp 
the eccentricity of the retained planet (i.e., $j=1$). In relatively
massive disks, the planets' eccentricity damping timescale may be 
shorter than disk depletion time scale while in disks with modest 
or low masses, planets may retain residual asymptotic eccentricity.
We estimate the local disk mass to be
\begin{equation}
M_{\rm disk} \sim \int^{2a}_{a/2} 2\pi r \Sigma_g dr \sim 
30 f_g \left(\frac{a}{1\mbox{AU}}\right) M_{\oplus},
\label{eq:mg_iso}
\end{equation}
where $\Sigma_g$ is disk gas surface density (the scaling factor $f_g$
is defined by Eq.~(\ref{eq:sigma_gas}) in \S 3.1), and $a$ 
is the semimajor axis of the remaining planet.
If the planet is beyond a truncation distance of the power-law disk,
$M_{\rm disk}$ is further reduced. In order to reflect the 
possibility of incomplete damping, we compare the angular momentum 
decrease due to the eccentricity damping with the total angular 
momentum of the local disk and limit the damping of eccentricity
by $\Delta e^2 = (M_{\rm disk}/m_1)^2$.  In our prescription, 
we impose a lower limit to the asymptotic eccentricity to be 
$e'_1 = \sqrt{e_1^2 - \Delta e^2}$ if $\Delta e < e_1$.

Since the energy carried by the ejected planet is negligibly 
small, the semimajor axis of the retained planet ($a_1$) is 
accurately obtained from the conservation of energy such that
\begin{equation}
\frac{m_1}{a_1} = \frac{m_1}{a_{1,0}} + \frac{m_2}{a_{2,0}}.
\label{eq:energy_consv}
\end{equation}
\end{enumerate}

\item[3-b)] {\it the non-ejection case:}
Both planets would be retained if both of their
$e_j^{\rm max} < 1$ after their close encounters.
In gas-free environment, such scatterings would 
recur until either one of the planet is ejected 
or they undergo direct collisions.  We assume 
this relaxation occurs on sufficiently short time
scale that the scattering events occur in the 
presence of disk gas.  Close scattering generally 
results in expansion of semimajor axis separation 
in addition to eccentricity excitation. The excited 
eccentricity is eventually damped by disk-planet 
interactions, leading the scattered planets to 
attain dynamical isolation.  The eccentricity 
damping is applied for both planets with the same 
prescription as that used in the ejection case.

\end{description}

\subsection*{A2. Three giants}

In \S 2.2, we discuss the interaction between three or more 
gas giant planets.  After the disk depletion, long-term secular 
perturbation between three or more gas giant planets leads
to dynamical instability and scattering between them.  In a 
gas-free environment, such scatterings recur until one or more
planet is ejected or some planets collide with each other.
We do not consider the possibility that after the onset
of orbit crossing, all members of multiple-planet systems 
(with three or more gas giants) can be retained.

N-body simulations \citep{Marzari02,Nagasawa08,Chatterjee08} 
showed that the most probable outcome of orbit crossing in 
systems of three gas giant planets is the retention of two
widely separated, eccentric planets and the ejection of one
planet.  The asymptotic orbits of the two retained planets
are stable within main-sequence lifetime of their host stars.

N-body simulations show that direct collisions may also occur
in relatively early stage before the eccentricities of all
three planets are fully excited.  We assume that $\sim 30\%$ 
of the systems undergo direct collisions.  Although there are
some uncertainties in the  collisional frequency,  we have found 
that the merger events do not significantly modify the planets'
asymptotic distribution provided their probability is less than 
$50\%$.  

The procedures to calculate the asymptotic eccentricities and 
semimajor axes of the retained planets are similar to those 
applied in the case of two giants except that we need to 
distinguish between the inwardly and outwardly scattered 
planet and evaluate their orbital elements independently.
The prescriptions for determining the dynamical outcomes are 
as follows:

\begin{description}
\item[1)] {\it Specification of the initial conditions:} 
In the population synthesis simulations, the criterion 
for orbital crossing that its evaluated time scale 
$\tau_{\rm cross}$ is smaller than the system evolution
time scale (see A3 below).  For a test case, we consider
systems of three planets with narrowly separated, nearly 
circular and coplanar initial orbits under the assumption 
that orbit crossing is initiated on a timescale of 
$\sim \tau_{\rm cross}$.  

\item[2)] {\it Determination of collision or scattering:} 
Based on the results of N body simulations, we assume 
direct collisions occur in $30\%$ of scattering events.
A merged planet is formed from a randomly selected 
pair of planets which are involved in the orbital crossings.
Its resultant semimajor axis and eccentricity are calculated 
under the assumed conservation of total mass, orbital energy 
(assuming that the energy dissipated during the collisions
equals to the binding energy of the colliding planets), and 
angular momentum.  In the N-body simulations, physical 
collisions between pairs of planets usually occur shortly 
after their orbits begin to cross, before their semimajor
axis and eccentricity have evolved significantly from their
original values. Therefore, we neglect the effect of 
eccentricity excitation induced by the orbital crossings.
In the remaining 70\% cases, we assume one or more planets
are ejected.  
\item[3)]  {\it Computation of trial eccentricities: }
First, we simulate the maximum eccentricities 
excited by the close encounters between multiple planets.
We denote planets 1 and 2 to be the most massive and second 
most massive planets. The maximum eccentricities excited through 
repeated close scatterings are calculated with 
Eqs.~(\ref{eq:e_escj0})  and (\ref{eq:e_escG}) where 
$e_{{\rm esc},12}$ is that of planet 1 and $e_{{\rm esc},j1}$ 
is that of the two less massive planets ($j=2,3$). Because all 
the planets cross each other's orbits, the eccentricities 
of planets 2 and 3 ($e_2$ and $e_3$) are primarily by the 
perturbation of planet 1 to the expected values of 
\begin{equation}
e_j^{\rm max} = 
\left\{
\begin{array}{ll}
\frac{m_1 {\cal R}_j}{(m_1 + m_j)}e_{\rm esc,1j} & ({\rm for} \;  j \ne 1), \\
\frac{m_2 {\cal R}_j}{(m_2 + m_1)} e_{\rm esc,12} & ({\rm for} \; j = 1).
\end{array}
\right.
\label{eq:e_3body_try}
\end{equation}
where ${\cal R}_j$ is a random number 
chosen from a Rayleigh distribution with the root mean 
square of unity (see Eq (\ref{eq:e_escj0}).

\item[4)]  {\it Identification of an ejected planet:} 
We identify the planet with the largest value of 
the maximum eccentricity. If its eccentricity exceeds
unity, we would consider this planet as an ejected body.

\item[5)] {\it Determination of the asymptotic eccentricities 
of retained planets:} 
In contrast to the two giants case, the orbital eccentricities 
of the two retained planets are not constrained by the amount
of residual angular momentum because it is freely exchanged and 
distributed between them.

Following the same argument as step 4 in Appendix A1,
we assume that the escaping planet ($j = {\rm ejc}$)
is ejected with a parabolic orbit, ie it carries little
total energy and its $e_{\rm ejc} \simeq 1$.

Assuming that the same degree of incomplete excitation, 
we obtain from Eq.~(\ref{eq:e_3body_try}) (with 
${\cal R}_j \simeq 1$) that
\begin{equation}
e_j  \sim  \frac{e_j^{\rm max}}{e_{\rm ejc}^{\rm max}} e_{\rm ejc} \sim 
\left\{
\begin{array}{ll}
\left(\frac{m_1 + m_{\rm ejc}}{m_1 + m_{j}}\right)^{1/2}
\left(\frac{R_1 + R_{\rm ejc}}{R_1 + R_{j}}\right)^{1/2}  
& ({\rm for} \;  j \ne 1), \\
\left(\frac{m_2}{m_1}\right)
\left(\frac{m_1 + m_{\rm ejc}}{m_1 + m_2}\right)^{1/2}
\left(\frac{R_1 + R_{\rm ejc}}{R_1 + R_2}\right)^{1/2}  
& ({\rm for} \; j = 1).
\end{array}
\right.
\label{eq:e_3body_ejc}
\end{equation}
Adding new values of ${\cal R}_j$ from those in Eq.~(\ref{eq:e_3body_try}),
we use
\begin{equation}
e_j  = 
\left\{
\begin{array}{ll}
\left(\frac{m_1 + m_{\rm ejc}}{m_1 + m_{j}}\right)^{1/2}
\left(\frac{R_1 + R_{\rm ejc}}{R_1 + R_{j}}\right)^{1/2} 
{\cal R}_j & ({\rm for} \;  j \ne 1), \\
\left(\frac{m_2}{m_1}\right)
\left(\frac{m_1 + m_{\rm ejc}}{m_1 + m_2}\right)^{1/2}
\left(\frac{R_1 + R_{\rm ejc}}{R_1 + R_2}\right)^{1/2} 
{\cal R}_j & ({\rm for} \; j = 1).
\end{array}
\right.
\label{eq:e_3body_ejc2}
\end{equation}

\item[6)] {\it Determination of an inwardly scattered planet:} 
The inwardly scattered planet is selected with a mass-square 
weighted probability.  We adopt this statistical weight because
the less massive member of the system tends to be scattered outward.
The mass weight function was calibrated with the N-body simulations. 

\item[7)] {\it Determination of the semimajor axis of an outwardly 
scattered body:} 
Since the outwardly scattered planet carries a small fraction of the
total energy, its asymptotic semimajor axis ($a_{\rm out}$)  is 
not well constrained by the constraint of total energy conservation.
The outer planet is scattered from the "vicinity" of the region
in which planets initially reside and its asymptotic periastron 
is close to the region.  We estimate the outer planet's periastron 
distance to be
\begin{equation}
a_{\rm out}(1-e_{\rm out}) = \sqrt{a_{\rm max} a_{\rm min}} 
+ a_{\rm max}{\cal R},
\label{eq:a_out}
\end{equation}
where $e_{\rm out}$ is its excited eccentricity of the outer planet,
and $a_{\rm max}$ and $a_{\rm min}$ are the maximum and minimum
semimajor axes of the planets in initial state prior to the orbit crossing.
Energy diffusion and semimajor axis redistribution generally occur 
before an ejection.  Due to the requirement of energy conservation, 
they generally lead to an expansion of the planetary system. 
Assuming that diffusion length is scaled by $a_{\rm max}$,
we adopt a "typical" semimajor axis of the outwardly scattered 
planet to be the right hand side of eq.~(\ref{eq:a_out}).
Since $e_{\rm out}$ is already determined in step 4,
$a_{\rm out}$ is given by Eq.~(\ref{eq:a_out}).

\item[8)] {\it Determination of the semimajor axis of an inwardly 
scattered body:} 
The semimajor axis of the inner planet ($a_{\rm in}$) is 
obtained under the constraint of energy conservation,
\begin{equation}
\frac{m_{\rm in}}{a_{\rm in}} = E - \frac{m_{\rm out}}{a_{\rm out}},
\label{eq:a_in}
\end{equation}
where $m_{\rm in}$ is the mass of the inner planet
and $E$ is the total energy calculated by the initial semimajor 
axes of the three planets.
Since the energy carried by the ejected body is very small,
$a_{\rm in} \simeq m_{\rm in}/E$. 
\end{description}

\subsection*{A3. The general case}

In the population synthesis calculations, giant planets and rocky/icy 
planetary embryos form in common disk environments, co-exist and 
dynamically evolve during and after the gas depletion. During their
early stages of their life span, i.e., while there is still substantial 
disk gas, we only consider scattering events between two planets
involving a gas giant when their orbital separation becomes less 
than $\Delta a_{\rm c}  \simeq 2\sqrt{3} r_{\rm H}$. Under this 
condition, we apply the prescription given in \S 2.1. 

The prescriptions for scattering after disk gas depletion in a general 
case are as follows:  
\begin{description}
\item[1)] 
We identify "giant planets" from a list of all the bodies in the system
by the conditions that (i) the planetary mass $m > 30 M_{\oplus}$ and 
(ii) $e_{\rm esc} > 1$ where 
\begin{equation}
e_{{\rm esc}} = \frac{v_{{\rm esc}}}{v_{\rm K}}
= \frac{\sqrt{2Gm/R}}{\sqrt{G M_{\ast}/a}}
\simeq 1.6 \left(\frac{m}{M_{\rm J}}\right)^{1/3}
  \left(\frac{\rho}{1{\rm gcm}^{-3}}\right)^{1/6}
  \left(\frac{a}{1{\rm AU}}\right)^{1/2}.
\label{eq:e_escG_cond}
\end{equation}

We evaluate $\tau_{\rm cross}$ for all pairs of the giant planets,
using the fitting formula obtained by \citet{Zhou07} with
some minor modifications:
\begin{equation}
\log \left(\frac{\tau_{\rm cross}}{T_{\rm K}} \right)
= A + B \log \left(\frac{b}{2.3 r_{\rm H}}\right),
\label{eq:tau_cross}
\end{equation}
where $T_{\rm K}$ is Keplerian time at the mean semimajor axis 
$a (= \sqrt{a_i a_j})$ of the pair, $b = | a_i - a_j |$, 
$r_{\rm H} = ((m_i+m_j)/3 M_{\ast})^{1/3} {\rm min}(a_i,a_j)$
(which is slightly modified from Paper VI), and
\begin{equation}
\begin{array}{l}
A = -2 + e_0 - 0.27 \log \mu, \\
B = 18.7 + 1.1 \log \mu - (16.8 + 1.2 \log \mu)e_0, \\
{\displaystyle e_0 = \frac{1}{2} \frac{(e_i + e_j)a}{b}}, \\
{\displaystyle \mu = \frac{(m_i + m_j)/2}{M_{\ast}}}.
\end{array}
\end{equation}

\item[2-a)]
We assume that the systems of only two giant planets are stable against
mutual long-term secular perturbations such as that between 
Jupiter and Saturn.  Nevertheless, systems of three or more gas 
giant planets with widely-separated nearly circular orbits may 
retain their initial semimajor axes and eccentricities.  But, 
if $\tau_{\rm cross}$ of some giant planet with two or more other giants
is less than the age of the system ($\tau_{\rm sys}$), we would assume 
that orbit crossing of these planets has occurred at 
$t = \tau_{\rm sys}+\tau_{\rm cross}$.  In our population 
synthesis models, if the expected $t$ is less than the 
calculation termination time that is $10^9$ years, we would
apply the following procedures to simulate the outcomes of 
orbital crossing among giant planets.

\begin{enumerate}
\item 
The pair of gas giant planets with the shortest orbital 
crossing time ($\tau^*_{\rm cross}$) is 
assumed to undergo close encounters before any other pairs.  
We select giant planets which participate in follow-up
(secondary) orbit crossings using the conditions that
radial excursions of their initial orbits overlap with 
the hypothetical orbits of the pair with $\tau^*_{\rm cross}$ 
predicted by the maximum eccentricities given by 
eq.~(\ref{eq:e_3body_try}).

\item We follow step 2 to 6 of three giants case in A2.
For systems with more than 3 gas giants, we assume only one 
gas giant planet is scattered inwardly while all the other
gas giant planets are scattered outwardly.  This assumption
is based on the results of N-body simulations which suggest
this is the most common outcome.

\item We calculate $\tau_{\rm cross}$ from the new orbital
configurations and go back to step 1 until 
$\tau_{\rm sys}+\tau^*_{\rm cross}$ exceeds $10^9$ years.
When the number of remaining gas giant planets is reduced
to two, we no longer apply this procedure.  
\end{enumerate}
Finally, we remove all the planets other than the gas giant 
planets.  This operation is based on the assumption
that violent secular perturbations from the highly
eccentric giants destabilize the other planets' orbits \citep{Matsumura13}.

\item[2-b)]
In systems without any gas giant planets or systems with widely
separated gas giants in which the orbit crossing time scale is 
longer than the age of their host stars, we consider the possibility
orbit crossing and dynamical relaxation among the small planets.
In these simulations, we assume negligible dynamical perturbation on
the gas giant planets by the smaller super-Earths and terrestrial planets.
The procedures for simulating the evolution of the smaller planets 
are extensively described in Paper VI, except for step 6
in which perturbations from a giant planet(s) are taken into account. 
Here we present a brief summary (for details, see Paper VI).

\begin{enumerate}
\item {\it Specification of the initial conditions.}
We select non-giant planets that can undergo eccentricity excitations.
We assume the amount of residual planetesimals is small enough that
dynamical friction from the planetesimals does not inhibits the
eccentricity excitations by mutual distant perturbations of the planets.

\item {\it Evaluation of $\tau_{\rm cross}$.}
We evaluate $\tau_{\rm cross}$ of all the low-mass planets, using
eq.~(\ref{eq:tau_cross}).  The pair of planets with the shortest orbit 
crossing time ($\tau^*_{\rm cross}$) is assumed to undergo close 
encounters before any other pairs.  

\item {\it Determination on the excited eccentricities and 
changes in semimajor axes of orbit-crossing planets.} 
We determine non-giant planets participating in the secondary orbit crossing 
and evaluate their excited eccentricities with eq.~(\ref{eq:e_3body_try}).
We determine changes in their semimajor axes due to radial diffusion by
successive close encounters after adjustment to satisfy
conservation of energy.

\item {\it Identification of collision pairs.}
Close encounters can also lead to physical collisions.  We 
assume all physical collisions are cohesive and they lead to 
merger events. Among the planets in the orbit crossing group, 
a colliding pair is chosen with a weighted 
collisional probability which is proportional to $a^{-3}$.

\item {\it Creation of a merger.}
A merged planet is created from the colliding pair. 
It acquires the total mass of the colliding pair. 
We determine orbital elements of the merged planet,
using conservation of energy and Laplace-Runge-Lenz vector.
We randomly select the relative angle between longitudes of 
periastrons of the colliding pair 
from its range that allows actual crossing of their unperturbed orbits. 

\item {\it Imposition of perturbations from the giant planets.}
If there is one or more gas giants in the system, the following 
procedure is applied.  If the range of radial excursion of 
non-giant planets in the new orbital configuration overlaps with 
the strongly-perturbed zone of any gas giant planets, we assume 
that the non-giant planets are scattered by the giant planet.
The width of the strongly-perturbed zone  is assumed to be 
$\sim 3.5 r_{\rm H}$.

Using eq.~(\ref{eq:e_3body_try}), we evaluate eccentricities of the
non-giant planets excited by the giant planets.  If the perturbed 
$e$ exceeds unity, we regard that the planet is ejected.  If it is 
less than unity, we evaluate a new semimajor axis for the scattered 
planet by $a_{\rm new} = a_{\rm old}/(1-e)$.  This procedure is 
added to the prescriptions described in Paper VI.
 
\item {\it Update of the system.}
We update $\tau_{\rm sys}$, adding $\tau_{\rm cross}$ and
a collision timescale after onset of orbit crossing. 
We go back to step 2 until
the updated $\tau_{\rm sys}$ exceeds $\sim 10^9$ years.
\end{enumerate}
\end{description}

{}

\end{document}